\documentclass[preprint,longabstract,8pt]{aastex}

\newcommand{\htw}{H$_2$}
\newcommand{\mhtw}{{\rm H_2}}
\newcommand{\mceio}{{\rm C^{18}O}}
\newcommand{\mthco}{{\rm ^{13}CO}}
\newcommand{\mco}{{\rm ^{12}CO}}
\newcommand{\coei}{C$^{18}$O}
\newcommand{\thco}{$^{13}$CO}
\newcommand{\co}{$^{12}$CO}
\newcommand{\ceio}{C$^{18}$O}
\newcommand{\av}{$A_{V}$}
\newcommand{\mav}{A_V}
\newcommand{\avo}{$A_{V13}$}
\newcommand{\mavo}{A_{V13}}

\newcommand{\sourcel}{Westend}

\newcommand{\cc}{ {\rm cm^{-3}} }
\newcommand{\kkm}{{\rm K\,km\, s^{-1}} }
\newcommand{\cmkm}{{\rm cm^{-2}\,K^{-1}\,km^{-1}\, s}}

\begin{document}

\title{CO isotopologues in the Perseus Molecular Cloud Complex: 
the $X$-factor and regional variations}
\shorttitle{CO isotopologues in Perseus}

\shortauthors{J. E. Pineda, P. Caselli and A. A. Goodman}
\author{Jaime E. Pineda}
\affil{Harvard-Smithsonian Center for Astrophysics, 60 Garden St., MS-10, Cambridge, MA 02138, USA}
\email{jpineda@cfa.harvard.edu}
\author{Paola Caselli}
\affil{School of Physics and Astronomy, University of Leeds, LS2 9JT, UK}
\affil{INAF, Osservatorio Astrofisico di Arcetri, Largo E. Fermi 5, I-50125 Firenze, Italy}
\email{p.caselli@leeds.ac.uk}
\and
\author{Alyssa A. Goodman}
\affil{Harvard-Smithsonian Center for Astrophysics, 60 Garden St., MS-42, Cambridge, MA 02138, USA}
\email{agoodman@cfa.harvard.edu}

\begin{abstract}
The COMPLETE Survey of Star-Forming regions offers an unusually comprehensive and diverse set of measurements of the distribution and temperature of dust and gas in molecular clouds, and in this paper we use those data to find new calibrations of the ``X-factor" and the \thco\ abundance within Perseus.  To carry out our analysis, we: 1) apply the NICER (Near-Infrared Color Excess Method Revisited) algorithm to 2MASS data to measure dust extinction; 2) use dust temperatures derived from re-processed IRAS data; and 3) make use of the \co\ and 
\thco~(1-0) transition maps gathered by COMPLETE to measure gas distribution and temperature. 
Here, we divide Perseus into six sub-regions, using groupings in a plot of dust 
temperature as a function of LSR velocity. 
The standard $X$ factor,  $X\equiv N(\mhtw)/W(\mco)$, is derived both for the whole 
Perseus Complex and for each of the six sub-regions with values consistent with 
previous estimates. However, the $X$ factor is heavily affected by the saturation 
of the emission above \av$\sim$4~mag, and variations are found between regions. 
We derive linear fits to relate $W(\mco)$ and \av\ using only points below 4~mag of extinction.
This linear fit yields a better estimation of the \av\ than the $X$ factor.
We derive linear relations of $W(\mthco), N(\mthco)$ and $W(\mceio)$ with \av.
In general, the extinction threshold above which \thco~(1-0) and \ceio~(1-0) are detected 
is about 1~mag larger than previous estimates, 
so that a more efficient shielding is needed for the formation of CO than previously thought.
The fractional abundances (w.r.t. \htw\ molecules) are in agreement with previous works. 
The (1-0) lines of \co\ and \thco\ saturate above 4 and 5~mag, respectively, whereas
\ceio~(1-0) never saturates in the whole \av\ range probed by our study (up to 10~mag).
Approximately 60\% of the positions with \co~(1-0) emission have sub-thermally 
excited lines, and almost all positions have \co~(1-0) excitation temperatures below 
the dust temperature. We compare our data with PDR models using the Meudon code, 
finding that \co~(1-0) and  \thco~(1-0) emission can be explained by these uniform 
slab models with densities ranging between about $10^3$ and $10^4~\cc$.  
In general, local variations in the volume density and non-thermal motions
(linked to different star formation activity) can explain the observations.
Higher densities
are needed to reproduce CO data toward active star forming sites, such as NGC~1333, 
where the larger internal motions driven by the young protostars allow more photons 
from the embedded high density cores to escape the cloud.  In the most quiescent 
region, B5, the \co\ and 
\thco\ emission appears to arise from an almost uniform thin layer of 
molecular material at densities around $10^4~\cc$, and the integrated 
intensities of the two CO isotopologues are the lowest in the whole complex.  
\end{abstract}


\section{Introduction}
Although \htw\ is the most abundant molecule in the interstellar 
medium (by about four orders of magnitude), it cannot be used as a 
tracer of the physical conditions in a molecular cloud. In fact, being a homonuclear 
species,  \htw\ does not have an electric dipole moment and even the 
lowest (electric quadrupole) rotational transitions require 
temperatures and densities well above those found in typical molecular 
clouds. The next most abundant molecule is \co, and since its 
discovery by \cite{Wilson:CO} it has been considered the best tracer 
of \htw\  and the total mass of molecular clouds 
\citep[e.g.][]{Combes:ARAA}.  Because of its relatively large 
abundance and excitation properties, the \co~(1-0) line is typically 
optically thick in molecular clouds, so that rarer CO isotopomers (in 
particular \thco) have also been used to trace the cloud mass in the 
most opaque regions. However, previous attempts to derive cloud 
masses from the thinner isotopomers present a large scatter (factor of 
10), indicating that \thco\ line strength is not an entirely straightforward indicator of gas column. 
In particular, the 
following points should be taken into account to derive less uncertain 
conversions: 1) preferential photodissociation of \thco \ at low optical 
extinctions, \av; 2) active chemical fractionation in the presence of 
$^{13}$C$^+$ ions, enhancing the relative abundance of \thco\ in cold 
gas \citep{Watson:1977}; and 3) variations in the \co /\thco\ abundance 
ratio across the Galaxy.

The conversion between \co~(1-0) integrated intensity ($W(\mco)$) and 
\htw\ column density ($N(\mhtw)$) is usually made using the so-called
``X-factor," defined as
\begin{equation}
X\equiv \frac{N(\mhtw)}{W(\mco)}~.
\end{equation}
In order to calibrate this ratio, one needs to measure the column of \htw\, which is difficult to do reliably. 

One method to derive this ratio uses only the \co\ emission and the assumption 
that molecular clouds are close to virial equilibrium. \cite{Solomon:relations} found 
a tight relation between the molecular cloud virial mass and the \co\ luminosity 
($M_{\rm VT}=39(L_{\rm CO})^{0.81}$). This relation enabled them to derive a ratio 
of ${\bf 3.0} \times 10^{20}~\cmkm$ for the median mass of the sample ($10^5~M_\odot$). 
In this case, the main source of uncertainty is the assumption of virial equilibrium for the 
molecular clouds.

Another method uses the generation of gamma rays by the collision of cosmic rays with 
H and \htw.  The \htw\ abundance is calculated using (assumed optically thin) 21-cm observations of H\ I and gamma ray observations in concert. \cite{Bloemen:1986} 
using COS-B data derived a value of $({\bf 2.8}\pm0.4) \times 10^{20}~\cmkm$
for the X factor, while \cite{Strong-Mattox:1996} derived $({\bf 1.9}\pm0.2) 
\times 10^{20}~\cmkm$ using the EGRET data. Uncertainties in gamma-ray-derived estimates of the X-factor stem from coarse resolution (0.5\degr\ and 0.2\degr\ for COS-B 
and EGRET, respectively) and from the assumption that there are no point 
sources of gamma rays.

An alternative method to derive the column density of \htw\ makes use of the 
IRAS 100~$\mu$m data to estimate the total column density of dust, which can 
then be used, assuming a constant dust-to-gas ratio (=0.01, value adopted also here), 
to estimate the total column density of Hydrogen 
($N({\rm H})=N(\textrm{\ion{H}{1}})+2N(\mhtw)$) and 
derive the conversion factor $X$. Using this method 
\cite{Dame:2001} derived a value of $({\bf 1.8}\pm 0.3)\times 10^{20}~\cmkm$ 
for the disk of the Milky Way, while \cite{deVries:Ursa} derived 
$({\bf 0.5}\pm 0.3)\times 10^{20}~\cmkm$ for the high-latitude far-infrared 
``cirrus'' clouds in Ursa Major. \cite{FLW82} find no correlation at 
all between $W(\mco)$ and $N(\mhtw)$ in Taurus (where the \co~(1-0) 
integrated intensity is roughly constant above 2 mag of visual 
extinction) while in $\rho$-Ophiuchus they derive $({\bf 1.8}\pm0.1)\times 
10^{20}~\cmkm$.

More recently, \cite{lombardi:pipe} studied the Pipe Cloud using an 
extinction map derived using the Near Infrared Color Excess Revisited 
(NICER) technique on 2MASS and \co~(1-0) data. They derived an $X$ 
factor of $({\bf 2.91}\pm 0.05) \times 10^{20}~\cmkm$, similar to what is 
found in cloud core regions and dark nebulae
\citep[see compilation by][]{Young:1982}. However, the factor derived 
by \cite{lombardi:pipe} includes a correction for Helium and uses the non-standard expression
$N({\rm H})=N(\textrm{\ion{H}{1}})+N(\mhtw)$, while using the standard 
definition of $N({\rm H})=N(\textrm{\ion{H}{1}})+2N(\mhtw)$ they would 
derive $({\bf 1.06}\pm 0.02)\times 10^{20}~\cmkm$.

\cite{FLW82} studied the relation between visual extinction and \thco\ 
data, and found that the number of \htw\ molecules per \thco\ molecule, 
[\htw/\thco], is ${\bf 3.7}\times 10^5$  in {\bf Taurus} and ${\bf 3.5}\times 10^5$ 
in {\bf Ophiuchus}, with extinction thresholds (below which no \thco\ is 
detected) of ${\bf 1.0}$ and ${\bf 1.6}$~mag, respectively. To make these 
estimates, \cite{FLW82} derived pencil-beam extinctions along the line of sight towards a 
handful of positions with background stars, by using near-infrared (NIR) 
spectroscopy from \cite{Elias:Oph} to estimate the spectral type and derive the 
extinction. 
However, pencil beam observations do not trace the same material probed by the 
molecular line emission observations, and this can introduce large uncertainties.
In fact, \cite{1999ApJ...517..264A} 
compared spectroscopically-determined extinction and IRAS-derived 
extinction in a stripe through Taurus, finding a 1-$\sigma$ dispersion  
of 15\% when the extinctions are compared between $0.9<\mav<3.0$~mag. 
This unavoidable dispersion is likely to affect previous column density 
derivations, such as \cite{FLW82}, as well.

Previous studies of the \thco\ abundance have also been carried out specifically in the portions of the
{\bf Perseus} Molecular Cloud Complex we study here. \cite{1986A&A...166..283B} used an 
extinction map derived from star counting on the Palomar Observatory 
Sky Survey \citep{1984A&AS...58..327C}. The angular resolution of 
their extinction map is $2.5\arcmin$, which is smoothed to $5\arcmin$ 
resolution for comparison with the $4.4\arcmin$ resolution of their 
\thco\ data, from which they derive an  [\htw/\thco] abundance and 
threshold extinction of ${\bf 3.8}\times 10^5$ and ${\bf 0.8}$~mag, respectively. 
\cite{1989ApJ...337..355L} compare \thco\  data of $2\arcmin$ resolution 
with the extinction map derived by \cite{1984A&AS...58..327C}, 
obtaining [\htw/\thco]$={\bf 3.6}\times 
10^5$ and an extinction threshold of ${\bf 0.5}$~mag.  However, this study was done only 
in the {\bf B5} region.  A summary of earlier measurements of the X-factor 
and \thco\ abundances can be found in \S\S~\ref{section:12co} and \ref{section:13co}.

The main disadvantage of previous calibrations done on Perseus is that 
they use {\it optically}-based star counting to create extinction maps of regions with high visual extinction
\citep{1986A&A...166..283B,1989ApJ...337..355L}. For Ophiuchus, \cite{scott-temp} 
find that the extinction derived using optical star counting \citep{Cambresy:1999} is systematically 
underestimated by $\sim 0.8$~mag in comparison with near-IR-based extinction mapping.  The offset (and resulting inaccuracy) is caused by the difficulty in fixing the zero-point extinction level -- in such a high extinction region -- in the optical star counting extinction map. In addition, the derived \av\ from 
optical star counting in Perseus does not have a large dynamic range 
($\mav<5~\rm{mag}$), while the NICER extinction map used in our work 
is accurate up to $10~\rm{mag}$ with small errors 
\citep[$\sigma_{\mav}<0.35~\rm{mag}$; see ][]{COMPLETE-I}.

The B5 region in Perseus is special, in that it has been the subject of an unusually high number of studies seeking to understand its basic physical properties.   As shown in Figure~\ref{regions}, B5 is somewhat isolated from the rest of Perseus, and it is only forming a very small number of stars, making it an obviously good choice for detailed study.
\cite{Young:B5} used a LVG model to derive an average density of 
$1.7\times10^3~\cc$ and kinetic temperatures $\sim 10-15$~K, 
but only for some stripes in the cloud. \cite{Bensch:2006} used 
\co\ and \thco\ maps with 
\ion{C}{1} pointing observations of 12 positions in a North-South stripe 
from the central B5 region to model the emission with a PDR code. From this analysis he 
derives average densities $\sim 3\times10^3-3\times10^4~\cc$.

The COMPLETE dataset offers the unique opportunity to study the 
emission of various CO isotopologues across the whole Perseus complex with 
unprecedented sensitivity and spatial resolution. In the present paper, 
COMPLETE data are analysed in detail to measure the \co\ excitation, 
the \thco\ abundance and the $X$ factor across the complex and 
study their variations. The observed changes in the measured quantities
are then related to local properties of the gas and dust. Using PDR codes, we 
find that local variations in the volume density 
and non-thermal motions (linked to different star formation activity) 
can explain the observations.

The extinction map, molecular and IRAS data used in this paper are 
presented in Sect.~\ref{sec-data}. The data selection is discussed in 
Sect.~\ref{sec-filter}. The six regions in which the Perseus Molecular Cloud
Complex is divided are identified in Sect.~\ref{identification}. 
Section \ref{sec-analysis} contains the analysis of the data, including the 
\thco\ column density determination and the curve of growth. Results can 
be found in Sect.~\ref{sec-results}.  The comparison between observations 
and PDR models is in Sec.~\ref{sec-pdr} and conclusions are listed in 
Sect.~\ref{sec-conclusions}.

\section{Data}\label{sec-data}
\subsection{Extinction Map}
We use the Near-Infrared Color Excess Method Revisited (NICER) technique \citep{NICER:Lombardi-Alves} on the Two
Micron All Sky Survey (2MASS) point source catalog to calculate K-band
extinctions.  To derive \av, we use the relation $A_K=0.112~\mav$
\citep{Rieke:1985}. The resulting extinction map has a fixed resolution of $5\arcmin$, and the
pixel scale is $2.5\arcmin$.  The total size of the map is $9\degr
\times 12\degr$ and is presented in \cite{COMPLETE-I} \citep[see][for more details]{pers-NIR}.

\begin{figure}
\centering
\epsscale{1.00}
\plotone{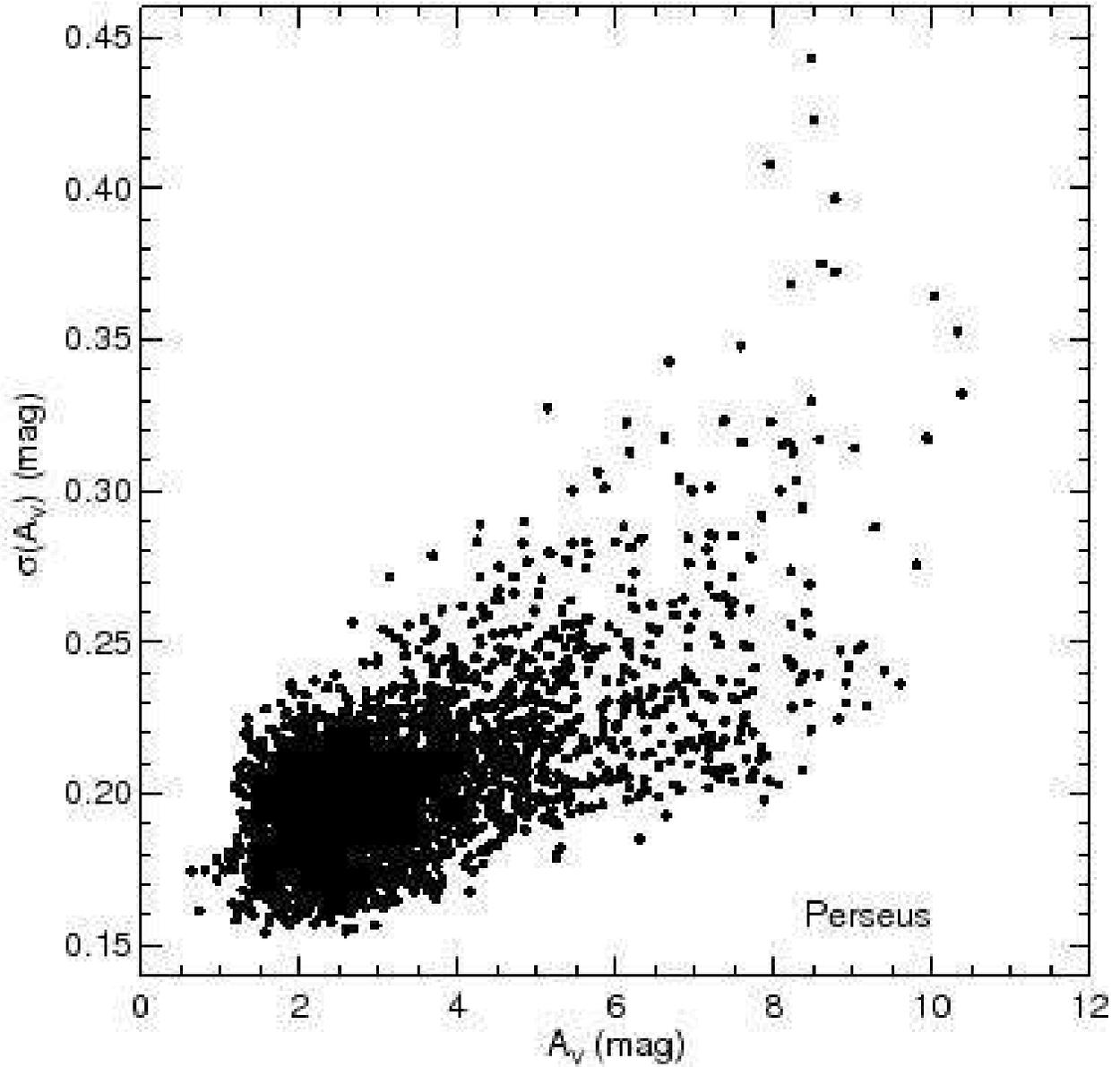}
\caption{\label{av-error} Error in the Perseus extinction map derived with 
NICER at each pixel versus the corresponding \av. Only the pixels used 
in this study are displayed here.}
\end{figure}

Figure~\ref{av-error} shows the estimated error (uncertainty) in the derived \av\ 
for all the points in the extinction map used in this study. The 
uncertainties are correlated with the extinction values because more stars per pixel allow for a more accurate measure of extinction.  Nevertheless, the linear slope in Figure~\ref{av-error} is smaller than $0.01$, so that the fractional uncertainty in {\it any} pixel's extinction measure remains quite small compared to its value. The median of the error for all of Perseus using NICER on 2MASS data is $0.2$~mag, while in the case 
of the extinction map derived by \cite{1984A&AS...58..327C} and used in 
\cite{1986A&A...166..283B,1989ApJ...337..355L} the typical error is 
$\sim 0.5~\rm{mag}$.  

While the improved accuracy offered by our new maps is significant, the increased dynamic range is even more critical to our analysis. The NICER extinction map of 
Perseus is accurate up to $\mav =10~\rm{mag}$, while the extinction map derived using star counting by \cite{1984A&AS...58..327C} dies out above $\sim 4-5~\rm{mag}$ of visual extinction.

\subsection{Molecular Data}\label{sec-data-mol}
We use the line maps of the COMPLETE Survey \citep{COMPLETE-I} to 
estimate \co~(1-0) and \thco~(1-0) column densities. Both lines 
were observed simultaneously using the FCRAO telescope. The line maps
cover an area of $\sim 6.25\degr \times 3\degr$ with a $46\arcsec$ 
beam in a $23\arcsec$ grid, and the positions included in our analysis 
are shown in Figure~\ref{regions}.
We correct the \co~(1-0) and 
\thco~(1-0) maps for a main-beam efficiency assumed to be $0.45$ and 
$0.49$, respectively 
(\url{http://www.astro.umass.edu/$\sim$fcrao/}). The flux calibration 
uncertainty is assumed to be 15\% (Mark Heyer, private communication).

In comparing the \co\ integrated intensity presented in 
\cite{Dame:2001} with the integrated intensity of our data smoothed to 
match the $1/8\,\degr$ beam resolution, 
we find the COMPLETE data and Dame's 
measurements to be well fitted by a linear relation of slope $0.9\pm0.1$ and 
off-set $-2\pm1~\kkm$.
We also compare the COMPLETE \thco\ integrated 
intensity with data from Bell Labs \citep{Padoan:Perseus}, after 
smoothing our data to the 100$\arcsec$ resolution and 1$\arcmin$ grid 
of Bell Labs data. These data are fitted by a linear relation of slope $0.986\pm0.003$ and 
off-set $-0.53\pm0.01~\kkm$, where the small deviation from the $1:1$ relation is most probably 
due to a small misalignment found between the images.

In addition, we use the \ceio~(1-0) data-cube presented by 
\cite{Hatchell:2003} \citep[and converted into FITS using CLASS90;][]{class90}, 
taken with FCRAO but with a smaller coverage and lower signal to noise. 

We convolve all data-cubes with a Gaussian beam to obtain the same 
$5\arcmin$ resolution as the NICER map, and then re-grid them to the 
extinction map grid of $2.5\arcmin$.

\subsection{Column Density and Dust Temperature from IRAS}
To estimate column density accurately from far-infrared (thermal) flux, one needs to also measure or calculate a temperature.  Normally, this is accomplished by making measurements at two separated far-IR wavelengths, making assumptions about dust emissivity, and then calculating two ``unknowns" (column density and temperature) from the two measurements of flux.  In the happy case where extinction mapping offers an independent measure of column density over a wide region, the conversion of FIR flux ratios to column density and temperature can be optimized so as to minimize point-to-point differences in comparisons of extinction- and emission-derived column density.  The column densities and temperatures derived from dust emission that we use in this paper and in \cite{alyssa-lognormal} come from the work of \cite{scott-temp}, who re-calibrated IRAS-based maps by using the 2MASS/NICER extinction maps discussed above to constrain the column density conversions.  

The ``IRAS" data used in \cite{scott-temp} come from the  the IRIS 
\citep[Improved Reprocessing of the IRAS Survey;][]{Miville05} flux 
maps at 60 and 100 \micron.  The IRIS data have better 
zodiacal light subtraction, calibration and zero level determinations, 
and destriping than the earlier ISSA IRAS survey release.

There are two important caveats to apply to the FIR-based column density and 
temperature measurements we use here (applicable to previous work as well).  
First, the {\it column densities} are only optimized to reduce scatter in a {\it global} (Perseus-wide) comparison of dust extinction and emission measures of column density -- they are still calculated based on the measured FIR fluxes at each point, and are thus not constrained to be identical to the 2MASS/NICER values.  Second, the dust {\it temperatures} derived by this method are uncertain due to: 1) unavoidable line-of-sight temperature variations, which cause increased scatter in comparisons with extinction-based measures and also cause a bias toward slightly higher temperatures \citep[see][]{Schnee:Bethell:Goodman}; and 2) the effect of emission from transiently heated Very Small Grains \citep[see][]{Schnee:spitzer}.  That said, the column density and temperature estimates based on \cite{scott-temp} and used in this paper do represent a dramatic improvement.

\section{Data  Editing for Analysis}\label{sec-filter}
The total number of pixels with data in all maps is 3765.
But in order to have high-quality data in every pixel of every map used in our analysis, we trim our maps to exclude particular positions where {\it any} data are not reliable.  The procedure used in data editing is described in this Section.

\subsection{Extinction Map}
Regions with both high stellar density and high extinction are typically associated with embedded populations of young stellar objects (YSOs).   Thus, in creating
extinction maps, including these regions, it would be foolish to assume 
that stars are background to the cloud and have ``typical'' near-IR colors.  Among YSOs, the more evolved objects observable in J, H, and K bands could in principle still be used to measure the extinction, but the fact that they are not 
background objects still produces an underestimation of the extinction. This 
affects the precision of the method much more than the non-stellar IR colors (infrared 
excess) of YSOs. Therefore, we exclude all the regions with 
a stellar density larger than $10$~stars per pixel from our analysis.  This criterion removes 34 pixels from our maps.  In addition to this editing, we exclude a box around each of the two main clusters in Perseus: IC348 and NGC1333, removing another 158 pixels from our data, as is evident in Figure~\ref{regions} (see Table~\ref{tab:cluster}). 

\subsection{Molecular Transitions}
To remain in the analysis, lines must have positive integrated intensities in both \co\ and \thco\ 
and peak brightness temperatures of at least 10 and 5 times the RMS noise for \co\ and \thco\ 
respectively.

Since \co\ is more abundant than \thco, and 
self-shielding is more effective for the \co~(1-0) transition than for the \thco~(1-0) transition,
\co\ emission is always more extended than \thco, both spatially and kinematically. In addition, \co\ lines are more affected  (broadened) by outflows and are optically-thicker than \thco.   Thus, line-widths for \co, $\sigma(\mco)$, should always be larger than those of \thco, $\sigma(\mthco)$. As a result, we keep only 
positions with \[\sigma(\mco)>0.8\,\sigma(\mthco)~,\] where the 0.8 
factor has been chosen to take into account the uncertainties in the 
line-width determination. The line-widths and central velocities of the 
spectra are obtained through Gaussian fits. This filtering accounts 
for only 72 out of the 324 pixels edited out in this study.

\subsection{Final Data Set}
The result of our editing leaves us with 3400 pixels (shown in Figure~\ref{regions}, see \S~\ref{identification} for details on color coding), out of an original 3724 pixels with \thco\ detections.  Note that our pixels are $2\times$ oversampled (see \S~\ref{sec-data-mol}), so that 3400 pixels amounts to 3400/4 independent measures.  

\section{Region Identification}
\label{identification}

\begin{figure}
\plotone{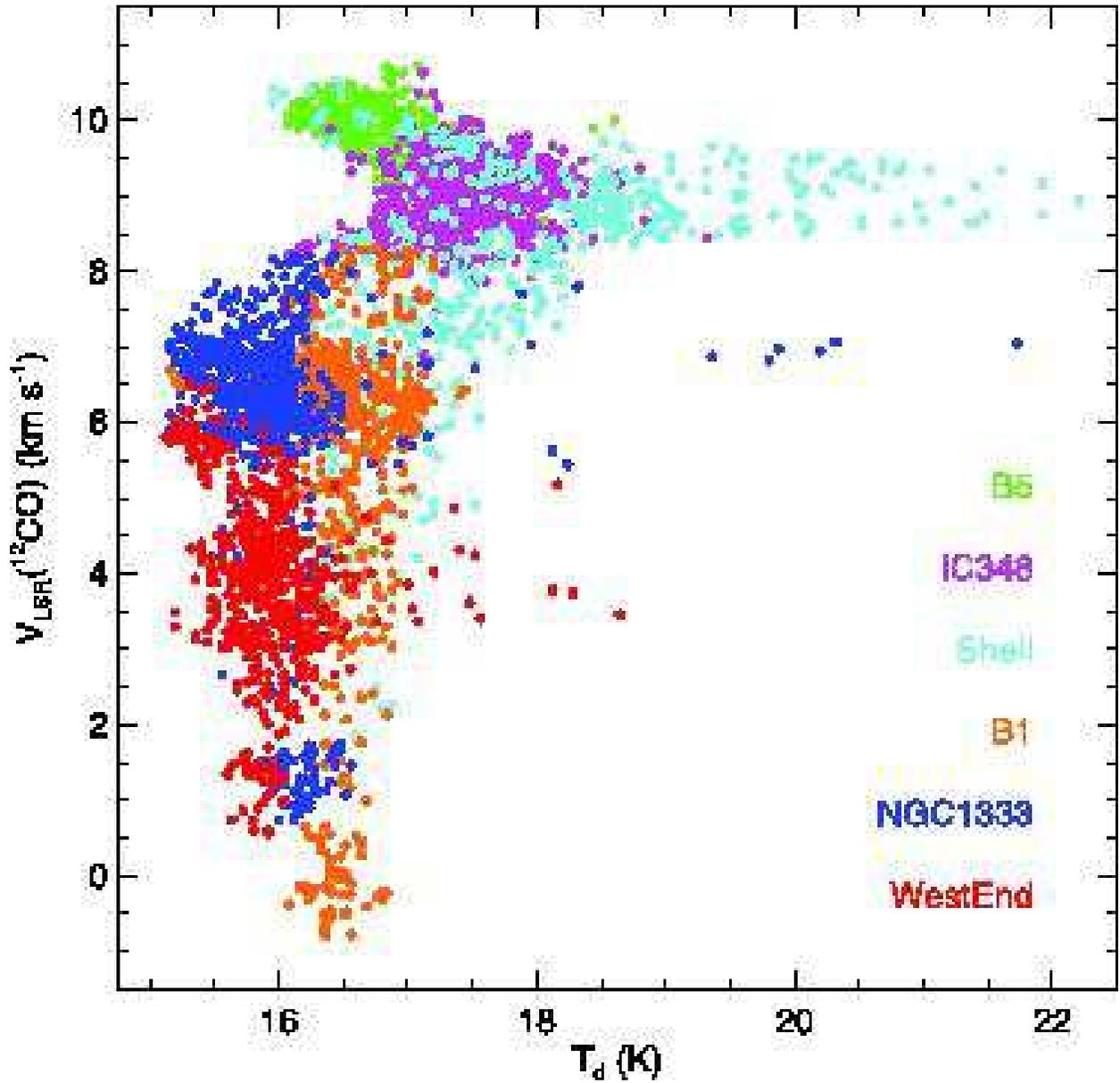}
\caption{\label{vcen-td} Central velocity compared with the dust 
temperature. Perseus is divided into six regions that are defined 
mainly in the space of physical parameters. This separation allows a 
better understanding of the cloud.}
\end{figure}

\begin{figure*}
\epsscale{0.67}
\plotone{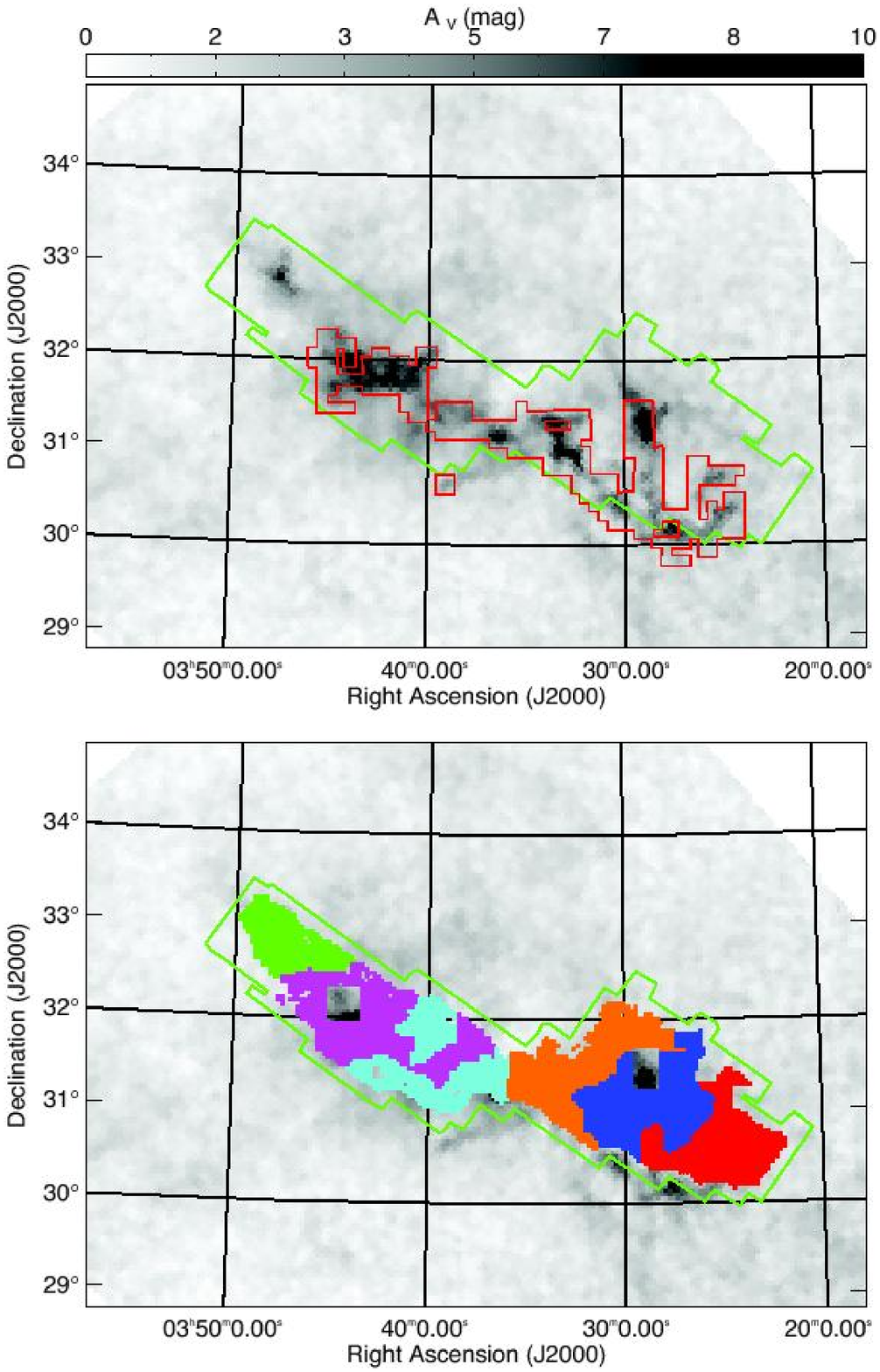}
\caption{({\it Top}) The extinction map derived using NICER. The 
green border is the region observed in \co\  and \thco~(1-0) by 
COMPLETE, while the red border is the region observed in \coei~(1-0) 
by \cite{Hatchell:2003}. ({\it Bottom}) The re-gridded molecular data that fulfil the 
requirements detailed in \S~\ref{sec-filter} are shown as 
boxes. Each of the defined regions are presented in a different 
color. B5 is green, IC~348 is magenta, the Shell is cyan, B1 is 
orange, NGC~1333 is blue and \sourcel\ is red. \label{regions} }
\end{figure*}

In trying to study Perseus as one object, it became apparent that much of the scatter in both the X-factor and in \thco\ abundance is caused by region-to-region variations (see \S\S~\ref{section:12co} and \ref{section:13co}).  So, we have divided the Perseus Molecular Cloud Complex into six regions 
with the help of 
several plots comparing different parameters (e.g. \co\ line-width, \co\ LSR velocity,
dust and excitation temperature). Figure~\ref{vcen-td} is an example of such plots, showing the \co\ 
velocity, $V_{LSR}$(\co) (the $V_{LSR}$ of \co\ and \thco\ are very similar)
as a function of dust temperature, $T_d$, for all the Perseus data. In 
this way, we avoid more arbitrary choices and minimize the data overlap 
among the different regions. As can be seen in Figure~\ref{vcen-td}, 
the six regions cluster around characteristic values of 
$V_{LSR}$ and $T_d$, so they are easily identified. 
Minor further refinements on the region
definitions is done to keep the regions physically connected, as shown in Figure~\ref{regions}. 
This step is needed primarily because there are regions with two velocity components along the 
line of sight and the Gaussian fitting can spontaneously switch from one 
component to the other. The effect of the two components is clearly 
seen in the points below 2.5~km\,s$^{-1}$ in Figure~\ref{vcen-td}, 
where points from three different geographical regions are merged into a single 
region of $V_{LSR}-T_d$ space. In 
Figure~\ref{regions} we show the final defined regions: B5, IC348, Shell, 
B1, NGC1333 and \sourcel. The Shell region is essentially the same as 
the shell--like feature discussed in \cite{2006ApJ...643..932R}. 
\sourcel\ encompasses L1448, L1455 and other dark clouds in the South-West 
part of Perseus. 
We would like to remark that the criteria adopted to identify the sub-regions 
in the Perseus Molecular Cloud have been chosen because they allow us to 
find (i) the minimum number of regions needed to improve the various 
correlations and (ii) the maximum number of regions with a statistically
significant number of data points.

\begin{figure}
\epsscale{0.60}
\plotone{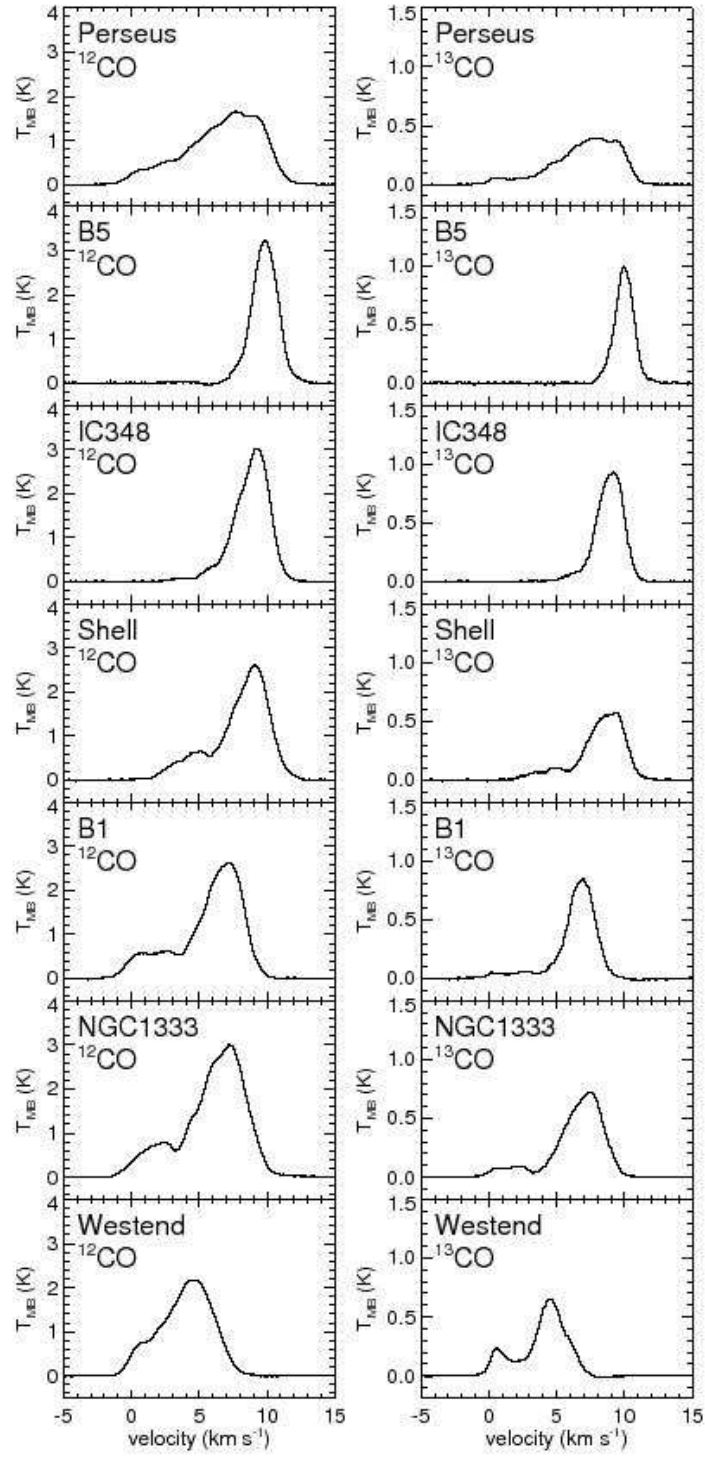}
\caption{ \label{aver-spec} Average \co~(1-0) and \thco~(1-0) spectra for all 
Perseus and the six sub-regions.}
\end{figure}

In Figure~\ref{aver-spec} we present the average \co\ and \thco\ 
spectra for the whole cloud and each region, while in 
Table~\ref{tab:prop} we show the main properties derived from the 
average spectra. The central velocity and velocity dispersion are 
computed from the average \thco\ with a Gaussian fit. We can see how 
different the region averages are from each other and the whole 
cloud. The gradient in central velocity across the cloud and the 
multiple components of the emission are clearly seen.

\section{Analysis}\label{sec-analysis}
\subsection{Column Density Determination}\label{sec-column}
To derive the \htw\  column density, $N(\mhtw)$, we assume that: (i) all the 
hydrogen traced by the derived extinction is in molecular form; (ii) the ratio 
between $N({\rm H})$ and $E(B-V)$ is $5.8\times 10^{21}~\rm{cm^{-2}\,mag^{-1}}$ 
as determined by \cite{bohlin:1978}; and (iii) $R_V=3.1$, to obtain 
\begin{equation}
\frac{N(\mhtw)}{\mav} = 9.4\times 10^{20}~\rm{cm^{-2}\ mag^{-1}}~.
\end{equation}
The ratio measured by \cite{bohlin:1978} was calculated only 
for low extinctions, and it is not clear whether extrapolating to higher extinction regions is wise.   In addition, it is well known that $R_V$ can 
increase up to values close to 4-6 in dense molecular clouds 
\citep[see e.g.][]{Draine:Review}, but we assume $R_V=3.1$ because it 
is the average value derived for the Milky Way and therefore our best estimation 
when this quantity has not been measured in the Perseus Cloud. This value of $R_V$ 
has also been used in all previous works, and therefore it facilitates the comparison.

To estimate \thco\ column densities we assume that the emission is 
optically thin and in Local Thermodynamic Equilibrium (LTE). To 
estimate the LTE column density, we have to assume an excitation 
temperature, $T_{ex}$, and optical depth, $\tau$.

In general, the intensity of an emission line, $I_{line}$, is 
\begin{equation}\label{eq:rt-1}
I_{line} = (S - I_0) (1-e^{-\tau}),
\end{equation}
where $S$ is the source function and $I_0$ is the initial impinging 
radiation field intensity. The radiation temperature is defined as 
\begin{equation}\label{def:T}
T_R = I_\nu \frac{c^2}{2\nu^2 k}~,
\end{equation}
where $I_\nu$  is the specific intensity and the filling factor is 
assumed to be unity. Assuming that the source function and initial 
intensity are black-bodies (with $I_\nu=B_\nu$) at $T_{ex}$ and $T_{bg}=2.7$~K, 
respectively, then we can write 
\begin{eqnarray}
T_{R} &=& 
T_0\left(\frac{1}{e^{T_0/T_{ex}}-1} - \frac{1}{e^{T_0/T_{bg}}-1}\right) (1-e^{-\tau}) ~, \label{eq:RT-gen}
\end{eqnarray}
where $T_0=h\nu/k$.

Assuming that the \co~(1-0) transition is optically thick, $\tau\rightarrow \infty$, and that 
$T_{max}(\mco)$ is the main beam brightness temperature at the peak of 
\co, we can derive the excitation 
temperature using eq.~(\ref{eq:RT-gen})
\begin{eqnarray}
T_{ex} &=& \frac{5.5~{\rm K}}{ \ln\left(1+ 5.5~{\rm K}/(T_{max}(\mco)+0.82~{\rm K})\right) } \label{tex} ,
\end{eqnarray}
where 5.5~K$\equiv h \nu(\mco) /k_{\rm B}$, with 
$\nu(\mco)$=115.3~GHz, the frequency of the \co~(1-0) line.   

Assuming that the excitation temperature of the \thco~(1-0) line is 
the same as for the \co~(1-0) line, the optical depth of \thco~(1-0) 
can be derived from eq.~(\ref{eq:RT-gen}),
\begin{eqnarray}
\tau(\mthco) &=& 
-\ln\left( 1-\frac{T_{max}(\mthco)/5.3~{\rm K}} 
{1/(e^{5.3~{\rm K}/T_{ex}}-1)-0.16}\right)~, \label{tau}
\end{eqnarray}
where $T_{max}(\mthco)$ is the main beam brightness temperature at the 
peak of \thco. 

The formal error for the excitation temperature, $\sigma(T_{ex})$, is 
\begin{equation}
\sigma(T_{ex}) =
\frac{\left[T_{ex}/\left(T_{max}(\mco)+0.82\right)\right]^2}
{1+5.5/\left(T_{max}(\mco)+0.82\right)}\,\sigma_{12}~,
\end{equation}
where $\sigma_{12}$ is the error in the \co\ peak temperature 
determination. We estimate that $\sigma_{12}$ is 
$0.15\,T_{max}(\mco)$ to account for the calibration uncertainty.

Using the definition of column density \citep{ToolsRA} and expressions (\ref{tex}) and (\ref{tau}), 
we derive the \thco\ column density as 
\begin{equation}
N(\mthco) = \left( \frac{\tau(\mthco)}{1-e^{-\tau(\mthco)}}\right) 3.0 \times 10^{14} 
\,\frac{W(\mthco)}{1-e^{-5.3/T_{ex}}} ~\rm{cm^{-2}}~,
\end{equation}
where the $W(\mthco)$ is the integrated intensity along the line of 
sight in units of ${\rm K\,km\,s^{-1}}$. This approximation is 
accurate to within $15\%$ for $\tau(\mthco)< 2$ \citep{spitzer:1968}, 
and always overestimates the column density for $\tau(\mthco) > 1$  \citep{spitzer:1968}.

In the  determination of the \thco\ column density, we use the derived 
excitation temperature instead of the dust temperature because the 
dust and gas are only coupled at volume densities above $\simeq 10^5~\cc$
 \citep[e.g.][]{Goldsmith:2001}, which are typically not 
traced by \co \ and \thco~(1-0) data.  Moreover, if the volume density 
of the gas falls below a few times $10^3~\cc$ (the critical 
density of the 1-0 transition), the \co\ lines are expected to be 
subthermally excited. 

\subsection{Curve of Growth}
Assuming LTE, the photon escape probability for a slab as 
a function of optical depth, $\beta(\tau)$, can be written as \citep{Tielens:book} 
\begin{equation}
\beta(\tau) =  \left\{
\begin{array}{ll}
	(1-\exp(-2.34\,\tau) )/4.68\,\tau   						& \tau<7\\
	1/4\tau \left[ \ln\left(\frac{\tau}{\sqrt{\pi}} \right) \right]^{1/2}  	& \tau>7
\end{array}~,\right.
\end{equation}
and the Doppler broadening parameter, $b$, is related to the atomic 
weight, $A$, and the gas temperature, $T$, by 
\begin{equation}
b = \sqrt{\frac{2\,k\,T}{m}} = 0.1290 \sqrt{\frac{T}{A}} ~\rm{km\,s^{-1}}~,
\end{equation}
where $m$ is the mass of the observed molecule.

The curve of growth relates the optical depth and the Doppler 
broadening parameter, $b$, with the integrated intensity, $W$, by
\begin{equation}
W=\int T_{MB}\,dv = T_R\,b\,f(\tau)~,
\end{equation}
where 
\begin{equation}
f(\tau) = 2\int_0^\tau \beta(\tilde{\tau})\,d\tilde{\tau} 
\end{equation}
in the Rayleigh-Jeans regime. In the above expressions $T_{MB}$ is the main beam 
brightness temperature (which is equal to $T_{R}$ for a filling factor of unity).

Assuming that \thco\ emission is optically thin and that the ratio 
between \thco\ and \co\ is constant in the region, we can write the 
optical depth as
\begin{equation}
\tau(\mco) = a\, W(\mthco)~,
\end{equation}
where $a$ is the conversion between $W(\mthco)$ and \co\ optical 
depth.

\section{Results} \label{sec-results}
\subsection{Curve of Growth Analysis}\label{sec-results-growth}
\begin{figure*}
\epsscale{1.0}
\plotone{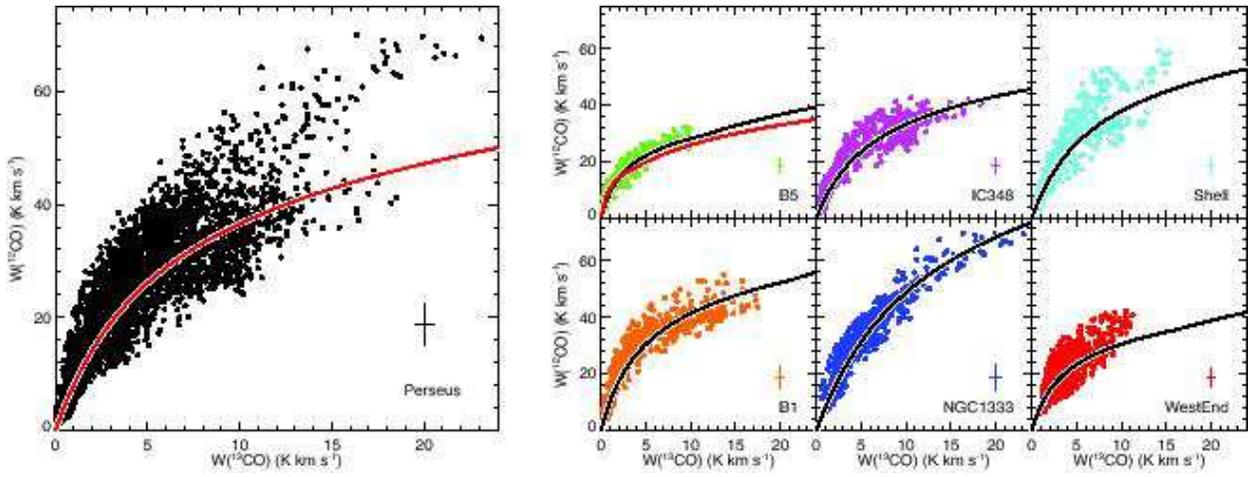}
\caption{\label{i1213} Integrated intensity of \co\ is plotted against 
the integrated intensity of \thco. The left panel shows all data used 
while right panel shows each region separately, using the same colors 
as in Figure~\ref{regions}. 
The median of the 1-sigma errors are shown in bottom right of each plot.
Solid lines are the growth curve fit, 
while the red curve in B5 is the fit from \cite{1989ApJ...337..355L}.}
\end{figure*}

As shown by \cite{1989ApJ...337..355L} in B5, \co\  and \thco\ 
integrated intensities are correlated and seem to be well described by 
the curve of growth \citep{spitzer:1978}. 
In Figure~\ref{i1213} we show \co\ and \thco\ integrated intensities 
for Perseus and the individual regions defined in \S~\ref{identification}. 
We perform a fit of the 
\co\ integrated intensity with a growth curve 
\begin{equation}
W(\mco) = T_R \,b \int_0^{aW(\mthco)}2\beta(\tilde{\tau})d\tilde{\tau}
\end{equation}
and present the fit results in Table~\ref{table-growth} and 
Figure~\ref{i1213}.

The fits for the growth curve are very good, considering the simplicity 
of the model. However, it is 
clear that the correlation is better in the individual regions than in 
the complex as a whole, with the exception of \sourcel, in which the 
fit is less good. Our results for B5 nicely agree both in shape and 
amplitude with \cite{1989ApJ...337..355L} (red solid line in 
Figure~\ref{i1213}). 

From the fit results we see that for gas at $12~{\rm K}$ (intermediate value between 
average excitation temperature and dust temperature) 
the derived Doppler parameters for B5, IC348, B1 and \sourcel\  are in reasonable 
agreement with $\sigma_V$ values listed in Table~\ref{tab:prop}. In 
NGC1333 and the Shell, a larger Doppler parameter is expected because 
the emission comes from multiple components, as seen in 
Figure~\ref{aver-spec}.

\subsection{The X--factor: using $\mco$ to derive \av}
\label{section:12co}
\begin{figure*}
\plotone{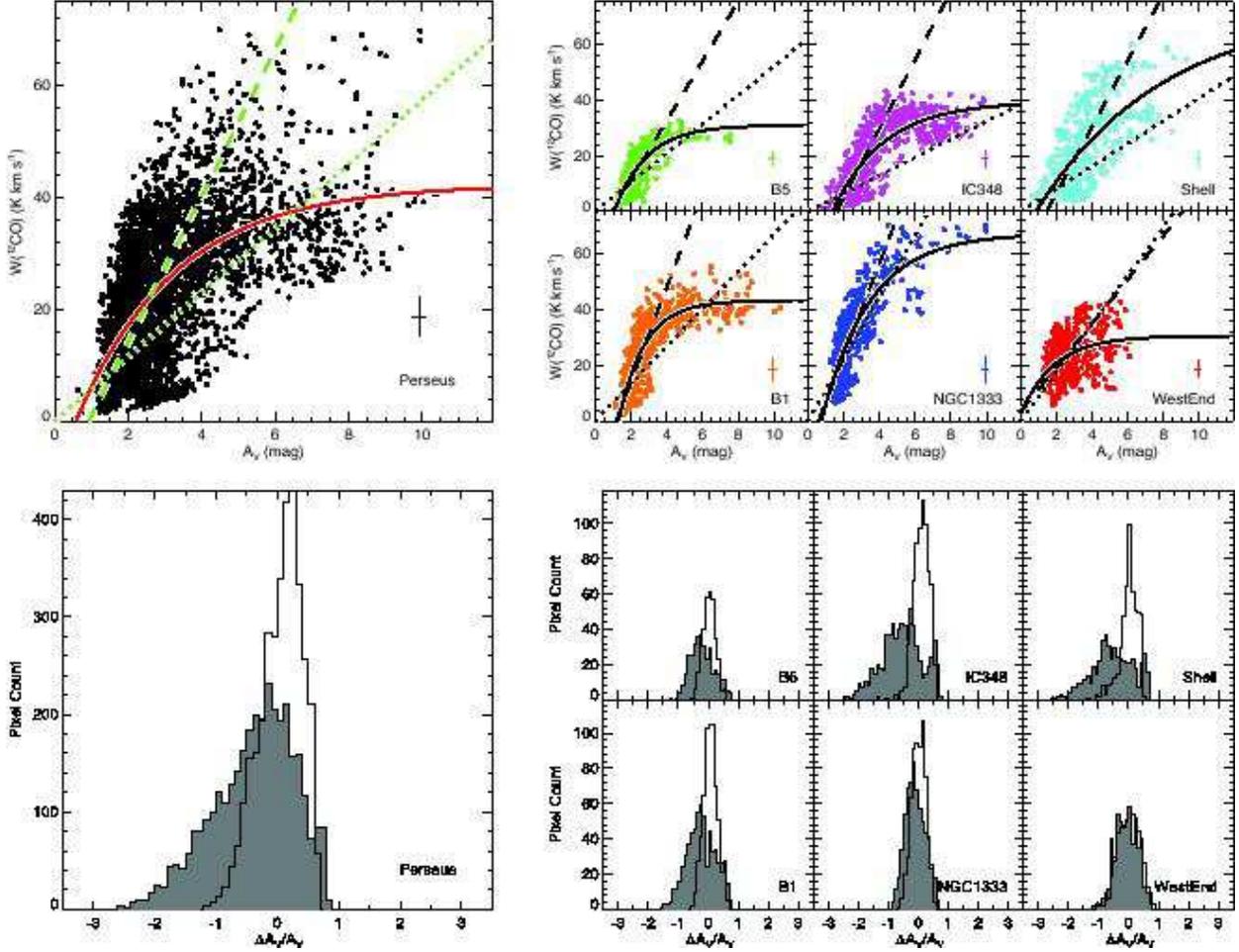}
\caption{\label{co-av} {\emph Top:} Integrated intensity of \co\ and visual extinction 
derived using NICER. The left panel shows all data used and the right 
panel shows each region separately, using the same colors as in 
Figure~\ref{regions}. Solid lines show the best fit of 
eq.~(\ref{co-func}), dotted and dashed lines are the standard 
$W(\mco)\,X=N(\mhtw)$ and straight line fits are only for points below 
\av=4, respectively.
{\emph Bottom:} Histograms of the difference between the observed and expected \av\ 
for  all data displayed in top panels. Filled and open histograms 
are for the standard $X$ factor and linear conversion, shown as dotted and dashed 
lines in upper panels, respectively.
}
\end{figure*}

The integrated intensity along the line of sight of the \co~(1-0)
transition, $W(\mco)$, is often used to trace the molecular material.
The conversion factor $X$, 
\begin{equation}
N(\mhtw) = X\, W(\mco),
\end{equation}
is derived, and to compare with previous results 
\citep[e.g.][]{Dame:2001} it is calculated as 
\begin{equation}
X=\left \langle \frac{9.4\times 10^{20}\,\mav }{W(\mco)} \right\rangle~.
\end{equation}
However, the \co\ emission saturates at $\mav\sim 4$~mag in every region, 
 as shown in Figure~\ref{co-av}. Therefore, we 
perform the estimation of the $X$ factor in two ways: (i) for all the 
points, and (ii) for only those points with $\mav < 4$~mag, where the column 
density can still be traced.

Figure~\ref{co-av} also shows that there is a threshold value of extinction below 
which no \co\ emission is detected.
To take this into account we fit the linear function 
\begin{equation}
\mav = A_{V12}+X_2 \frac{W(\mco)}{9.4\times 10^{20}},
\end{equation}
where $A_{V12}$ is the minimum extinction below which there is no \co\ 
emission and $X_2$ is the slope of the conversion (comparable to the 
$X$ factor). The linear fit is performed using the bivariate 
correlated errors and intrinsic scatter estimator (BCES), which takes 
into account errors in both axes and provides the least biased 
estimation of  the slopes \citep{1996ApJ...470..706A}. The results of 
the $X$ factor and the linear fits are presented in 
Table~\ref{table-co-X}.  In Figure~\ref{co-av}, we show only the results 
for points with $\mav < 4$~mag: a dotted line for the standard $X$ 
factor and a dashed line for the linear fit.

The values derived for $X$ using all the points (as previously done in 
the galactic determinations of $X$) are in agreement with the mean 
value of $(1.8\pm0.3)\times10^{20}~\cmkm$ derived by \cite{Dame:2001} 
for the Milky Way. However, this fit is not good in the unsaturated regime 
($\mav < 4$~mag). Performing a 
linear fit to all of the data, including the saturated emission, can give
unreasonable solutions, such as a negative minimum extinction needed to 
produce \co~(1-0) integrated intensity.

The linear fit performed to positions with $\mav < 4$~mag gives the 
best estimate for the extinction in the unsaturated regions but only 
provides a lower limit extinction estimate for the saturated regimes, 
while the standard $X$ factor provides a poor description of the data 
in both saturated and unsaturated regimes.
The histograms of the errors associated with the conversions derived for positions
with $\mav < 4$~mag (bottom panels of Figure~\ref{co-av}) show that the linear fit 
(open histogram) provides a more unbiased estimate of the extinction than 
the $X$ factor (filled histogram).
This improvement in the precision of the extinction estimate goes along a 
reduction in the errors. The width of a Gaussian fitted to the histograms of the 
regions implies a typical error of 40\% and 25\% for the X factor and the linear fit, 
respectively. Performing the same analysis on the histograms of the whole cloud 
gives an error of 59\% and 38\% for the X factor and the linear fit, respectively.

However, as we can see from Figure~\ref{co-av} the linear relation is 
not very accurate for $\mav > 4$~mag. Therefore, following the 
simplest solution of radiative transfer 
\[ I=I_0 \left(1-e^{-\tau}\right)~,\]
we fit it to our data with 
\begin{equation}\label{co-func}
W(\mco)=W_0\left(1-e^{-k(\mav-A_{k12})}\right)~,
\end{equation}
where $W_0$ is the integrated intensity at saturation, $A_{k12}$ is 
the minimum extinction needed to get \co\ emission, and $k$ is the 
conversion factor between the amount of extinction and the 
optical depth. We perform an unweighted fit of the 
non-linear function 
to the data, which yield solutions that better follow the overall shape
of the $W(\mco)$ as a function of \av. Unfortunately, the best fits for \sourcel\ and Perseus 
are quite poor and should be regarded with caution. The best parameters are listed 
in Table~\ref{table-co} and are shown in Figure~\ref{co-av} as solid curves.
We note that the threshold extinction shows a large scatter, even in the more 
accurate non-linear fit. This suggests that different environmental conditions 
are causing the observed scatter (see \S~\ref{sec-pdr}).

\subsection{Using $\mthco$ to derive $\mhtw$\ column densities}
\label{section:13co}
\begin{figure*}
\plotone{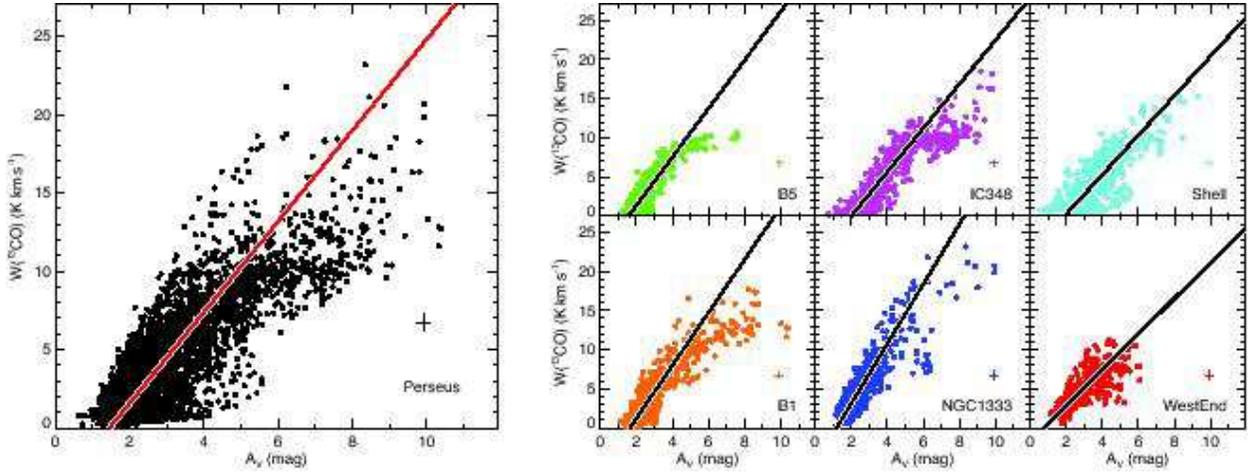}
\caption{\label{i13-av} Left panel shows \thco\  Integrated intensity 
compared with the extinction. The median of 1-sigma errors are shown 
in the bottom right corner of each plot. The best linear fit is shown 
with a solid line. The right panel shows same plots as the left panel, but 
for the individual regions.}
\end{figure*}

We expect the \thco~(1-0) transition to be optically thin at low 
extinction. If this is the case, a simple linear relation between the 
integrated intensity of \thco\ and \av\ should fit the data,
\begin{equation}
\mav = W(\mthco)\, B_{13} +A_{W13}~.
\end{equation}
The fit is done using the BCES algorithm for points with \av\ $<
5$~mag because there is some saturation in the emission. The results 
are presented in Table~\ref{table-w13} and in Figure~\ref{i13-av}. 
Once again, we find that the minimum extinction needed to detect 
\thco\ and the slope of the linear relation changes between the regions, 
though the variation is smaller than with the \co\ data. From 
Figure~\ref{i13-av} we can clearly see the effect of saturation around 
5~mag of extinction in IC348 and B1, similar to what was reported by 
\cite{Lada:1994} in the more distant IC~5146, with comparable linear 
resolution. When comparing our 
results with the linear fit derived by \cite{Lada:1994} we see that 
the fit parameters are quite different, suggesting that the IC~5146 
cloud is quite different from Perseus. The linear fit for IC~5146 
would indicate that for a region without extinction there is molecular 
emission, suggesting that the linear fit has been affected by points 
with saturated emission or that the fit errors are largely 
underestimated. Comparing our result for B5 with 
\cite{1989ApJ...337..355L}, we find that the slopes are slightly 
different, but this can be reconciled by performing the linear fit 
taking all the points ($B_{13}=0.37\pm0.1$, $A_{W13}=1.45\pm0.04$) as 
done by \cite{1989ApJ...337..355L}. However, 
the fit performed by \cite{1989ApJ...337..355L} was done using the 
extinction map derived by \cite{1986A&A...166..283B}, that is limited by \av\ 
values below 5~mag of visual extinction (due to a lack of detectable
background stars). Also, given that
their molecular data has a better resolution (1.5\arcmin) than the 
extinction map (2.5\arcmin), they interpolated the \thco\ to the extinction
positions instead of smoothing the data to the same resolution. 
Finally, the threshold extinction value differs from previous 
measurements mainly because it is hard to accurately define the zero 
point for extinction maps derived from optical star counting 
(as already mentioned, \citealt{scott-temp} reported a difference of $\sim0.8$~mag).

\begin{figure*}
\plotone{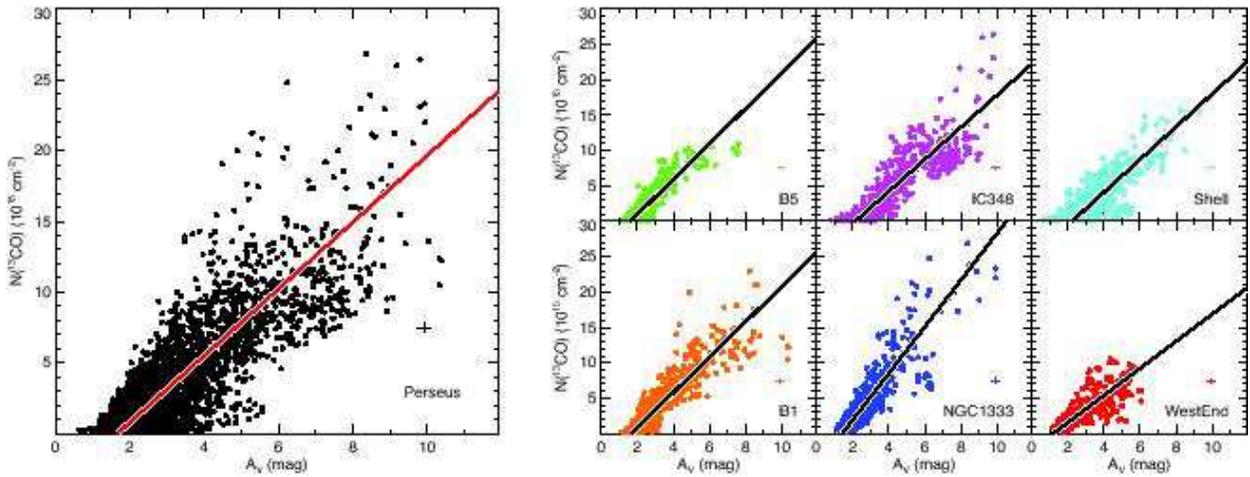}
\caption{\label{13co-av} The left panel shows \thco\ column densities as a function of visual extinction. 
The median of the 1-sigma errors are shown in 
bottom right of each plot. The excitation temperature is estimated 
using the \co\  peak temperature, assuming this transition to be optically 
thick. The best fit is shown with a solid line. The right panel show 
the same as the left panel, but separated into different regions.}
\end{figure*}

Following the column density determination shown in \S~\ref{sec-column}, which 
takes into account the effect of optical depth and excitation temperature, 
we investigate the relation between $N(\mthco)$ and \av, fitting a linear 
function to the data,
\begin{equation}
\mav =
N(\mthco)\, c +\mavo,
\end{equation}
where c and \avo\ are the parameters of the fit. From the fit we can derive 
the ratio of abundances between \htw\ and \thco\ as
\begin{equation}
\left[\frac{\mhtw}{\mthco}\right] = 9.4\times 10^{20}\ c~.
\end{equation}
The result for the 
fit over the whole cloud and by regions is shown in 
Figure~\ref{13co-av}, and in Table~\ref{table-cal} we present the best 
fit parameters. When performing the fit we estimate the error 
associated with the \thco\ column density as 15\%, due to the 
uncertainty in the calibration.

As shown in Figure~\ref{13co-av}, this relation presents a larger 
scatter, which is also present in individual regions \citep[and it is 
consistent with previous work; see e.g.][]{Combes:ARAA}.  As already 
pointed out, selective photodissociation and/or chemical fractionation 
can alter the simple linear relation at low \av\ values, whereas 
optical depth may start to be large at high \av\ (see e.g. the tendency 
of $N(\mthco )$ to flatten out at $\mav >5$~mag in B5, IC348 and B1).

The \thco\ abundances derived from the fit present significant 
variations from region to region (see Table~\ref{table-cal}). We do 
not find a correlation between the abundance and the threshold 
\avo. This suggests that {\it $\mthco$ abundance variations are mainly due 
to different chemical/physical properties in the inner regions of the 
cloud at $\mav > \mavo$}. 

Comparing our results with those in Table~\ref{table-cal}, we see that 
the abundances agree very well, within the errors, with previous values 
reported in Perseus, taking also into account  the $\sim 0.8$~mag difference 
found by \cite{scott-temp}. This difference is 
produced by the difficulty in defining the zero level of extinction, 
which is harder in the optical star counting method than using NICER.
The extinction threshold derived for Perseus is close to the one 
derived for $\rho$-Oph by \cite{FLW82}, but they do not have any data 
at \av\ below $2.5$~mag and their determination has been done only for 
points with $\mav>4$~mag.

It is important to note that the numbers cited from previous works do not include 
the 10-20\% calibration uncertainty that we do include in our results, and therefore 
our results are more accurate than previous ones.

\subsection{Using $\mceio$ to derive $\mhtw$\ column densities}
\label{section:c18o}
We fit a linear relation between the 
integrated intensity of \coei\ and \av,
\begin{equation}
\mav = W(\mceio)\, B_{18} +A_{W18}~.
\end{equation}

\begin{figure*}
\plotone{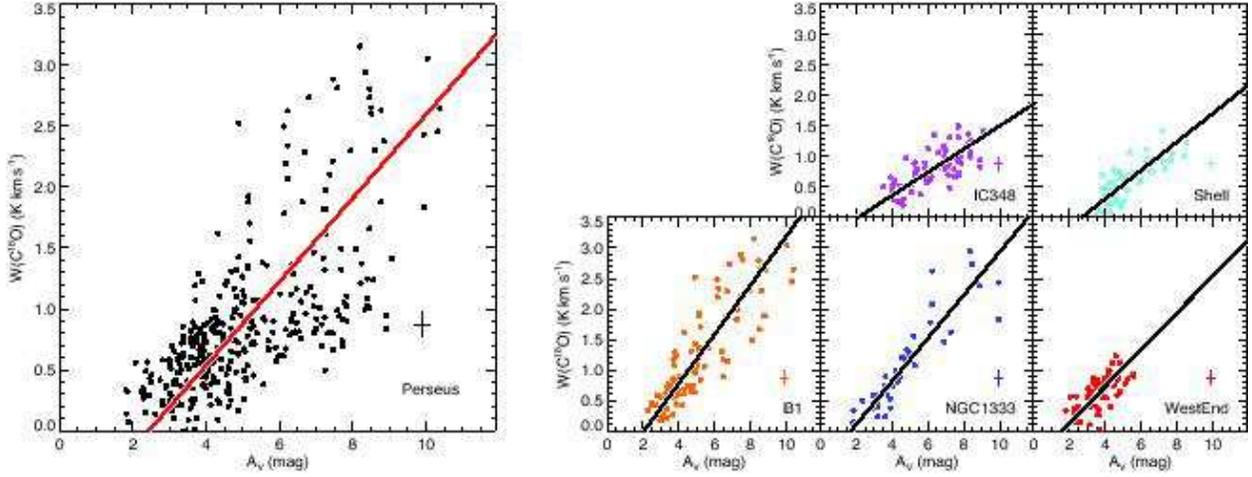}
\caption{\label{i18-av} The left panel shows \coei\ integrated intensities as a function of  
visual extinction. The median of 1-sigma errors are shown 
in the bottom right corner of each plot. The best linear fit is shown 
with a solid line. The right panel shows the same as the left panel, but 
separated into different regions.}
\end{figure*}

The fit results are presented in Table~\ref{table-w18} and in Figure~\ref{i18-av}. 
We find that, as in the case of \co\ and \thco, both parameters (extinction threshold 
and slope) vary between regions. However, due to the fewer points available per 
region, the errors are larger than for previous fits.

Despite the small number statistics, we still can see that \ceio\ is fairly linear 
up to at least 10~mag. When comparing our results with the linear fit derived 
by \cite{Lada:1994}, we see that the fit parameters are quite different. However, 
they performed the fit over a wider \av\ range (up to 15~mag), and it is possible that 
despite their efforts the fit could have been affected by emission with higher optical
depth.

\cite{1989ApJ...337..355L} derived a fit in B5. Unfortunately, we don't have 
coverage for B5, and therefore no direct comparison can be done. 
Nevertheless, we find that the threshold extinction value derived for B5 is systematically  
lower than the values derived here for other regions (similar to what is found in \thco) while 
the slope is in agreement (within the error bars) with the parameters derived here.

\subsection{$\mco~(1-0)$ Excitation Temperature vs. Extinction}\label{sec-tex}
Being collisionally excited, the excitation temperature of \co~(1-0) is 
expected to increase as we move from the outskirts of the cloud, where 
the extinction and volume densities may be lower than the critical 
density for the \co~(1-0) transition, to the most extinguished and 
densest regions, where \co~(1-0) is in LTE and faithfully traces the 
gas kinetic temperature. 

\begin{figure*}
\plotone{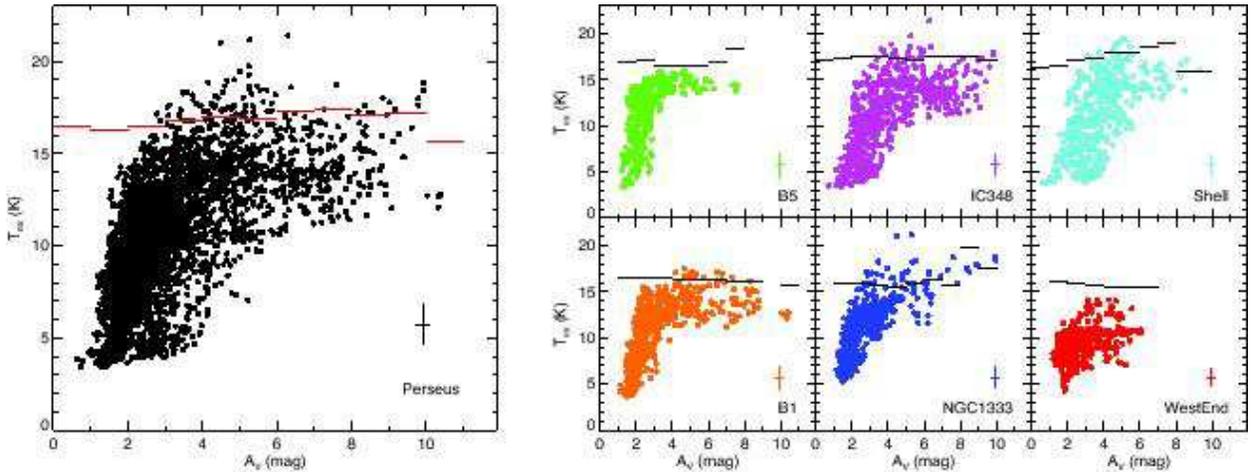}
\caption{\label{tex-av} Excitation temperature derived using \co\ 
plotted against the visual extinction at the same position. The
median of the 1-sigma errors are shown in bottom right of each plot.
The solid line is the median dust temperature for each sample within 
the extinction bin. The left panel shows all data used while the 
right panel shows them separated by regions.}
\end{figure*}

The excitation temperature, derived using \co\  and eq.~\ref{tex}, is 
shown as a function of the visual extinction in Figure~\ref{tex-av}.
In addition we plot the median dust temperature computed in bins of 
extinction as horizontal lines. When all the points are plotted (left 
panel of Figure~\ref{tex-av}) there is a poor correlation between 
excitation temperature and extinction. This is a direct result of 
mixing very different environments within the cloud in one plot. In 
fact, from the right panel of Figure~\ref{tex-av}, where the 
excitation temperature and visual extinction are plotted for 
individual regions, it is clear that the scatter is significantly 
lower and the excitation temperature rises from $\sim 5$~K at low 
\av\ ($\sim 2$~mag) up to a temperature close to the derived dust 
temperature in positions with $\mav>4$. The 
more quiescent regions (B1 and B5) present a smaller dispersion, 
whereas more active regions (IC348 and NGC1333) present a larger 
spread in the excitation temperature, probably because of the larger variation
of physical conditions along the line of sight and/or the multiple velocity components. 
The region labeled as 
``\sourcel'' has a very low excitation temperature when compared with 
the rest of the cloud. 
As shown in the right panel, in individual 
regions the dust temperature does not change more than approximately 
one degree except in the ``Shell'' which shows a steady increase with \av, 
and in NGC1333 where a peak of the dust temperature is present at 
$\mav\sim 8$~mag, probably due to the internal heating produced by the 
nearby embedded cluster. On the other hand, the \co\ excitation temperature 
ranges between 5 and 20~K.

It is important to note that almost all the points lie below the 
median dust temperature of the region, indicating that 
\co\ is tracing gas at volume densities well below 
$10^5~\cc$, the lower limit to have dust and gas coupling 
\citep{Goldsmith:2001}. 
The average excitation temperature for points above 4~mag is $13.8$~K, while the 
standard deviation is $2.3$~K. 
We count the number of positions where the \co\ emission is sub-thermal ($T_{ex}<11.5$~K)
obtaining that it is $\sim 60$~\%.
\sourcel\ is a region with \co\ excitation temperature 
always below the dust temperature. This could be due to a lower fraction of 
high density material compared to the other regions. 
It is interesting that in the regions in the North-East part of the cloud (B5, IC348 
and Shell) the dust temperature is closer to 
17~K while in the South-West part (B1, NGC1333 and \sourcel) it is closer to 16~K, 
suggesting variations in the ISRF across the Perseus Complex.

\section{Modeling using PDR code}\label{sec-pdr}
To relate the observed variations in the \co\ and \thco\ lines 
with changes in the physical properties of the regions, we 
use the Meudon PDR code \citep{Meudon:PDR}\footnote{Available through 
\url{http://aristote.obspm.fr/MIS/}}. This code includes most 
physical effects by explicit calculation; 
in particular it calculates the \co\ shielding, unlike the majority of 
other codes available \citep[e.g.][]{PDR:comparison},
where fitting formulae are used instead.

We use the abundances derived by \cite{Lee:1998} (see 
table~\ref{table:abun}) for clouds with high metal abundances (more 
appropriate for the material traced by \co\ and \thco) a $^{12}$C/$^{13}$C 
abundance ratio of 80 and a cosmic-ray ionization 
rate of $\zeta=10^{-17}~\rm{s^{-1}}$ (a $\zeta$ value 6 times larger than the
adopted one has also been considered, but found not to change the results 
by more than 10\%; see Appendix). To reproduce the observed [\htw/\thco] 
ratio we increased the $^{12}$C abundance by a factor of 1.8 compared to 
\cite{Lee:1998}. To create curves of integrated 
intensity as a function of \av, we run PDR models with different 
extinctions (\av$=0.5,1,1.5,2,2.5,3,4,5,8,10$\,mag). Each PDR 
calculation is performed over a grid of parameters, assuming a slab of 
constant density illuminated on both sides: (1) The turbulent velocity 
is fixed to the Doppler parameter derived from the curve of growth fit 
assuming $T_R=12$~K for each region (see Table~\ref{table-growth} and 
\S~\ref{sec-results-growth}). 
(2) The volume density 
$n=10^3,5\times10^3,10^4,5\times10^4~\cc$. (3) The radiation field is 
$\chi=1\textrm{ and }3$ times the standard Draine's radiation 
field \citep{Draine:1978}  (other $\chi$ values have been explored and reported in the 
Appendix).

\begin{figure}
\epsscale{0.5}
\plotone{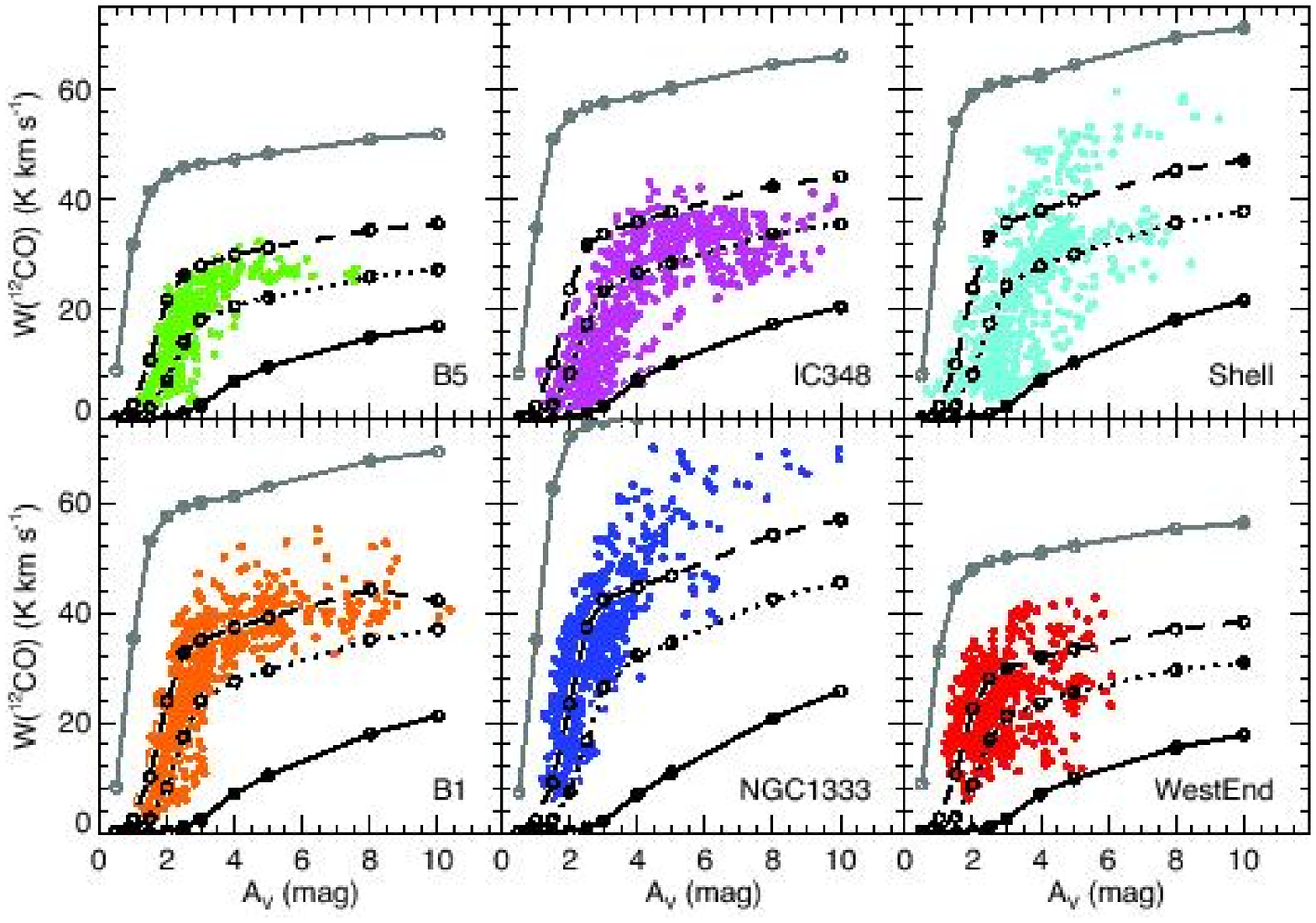}
\plotone{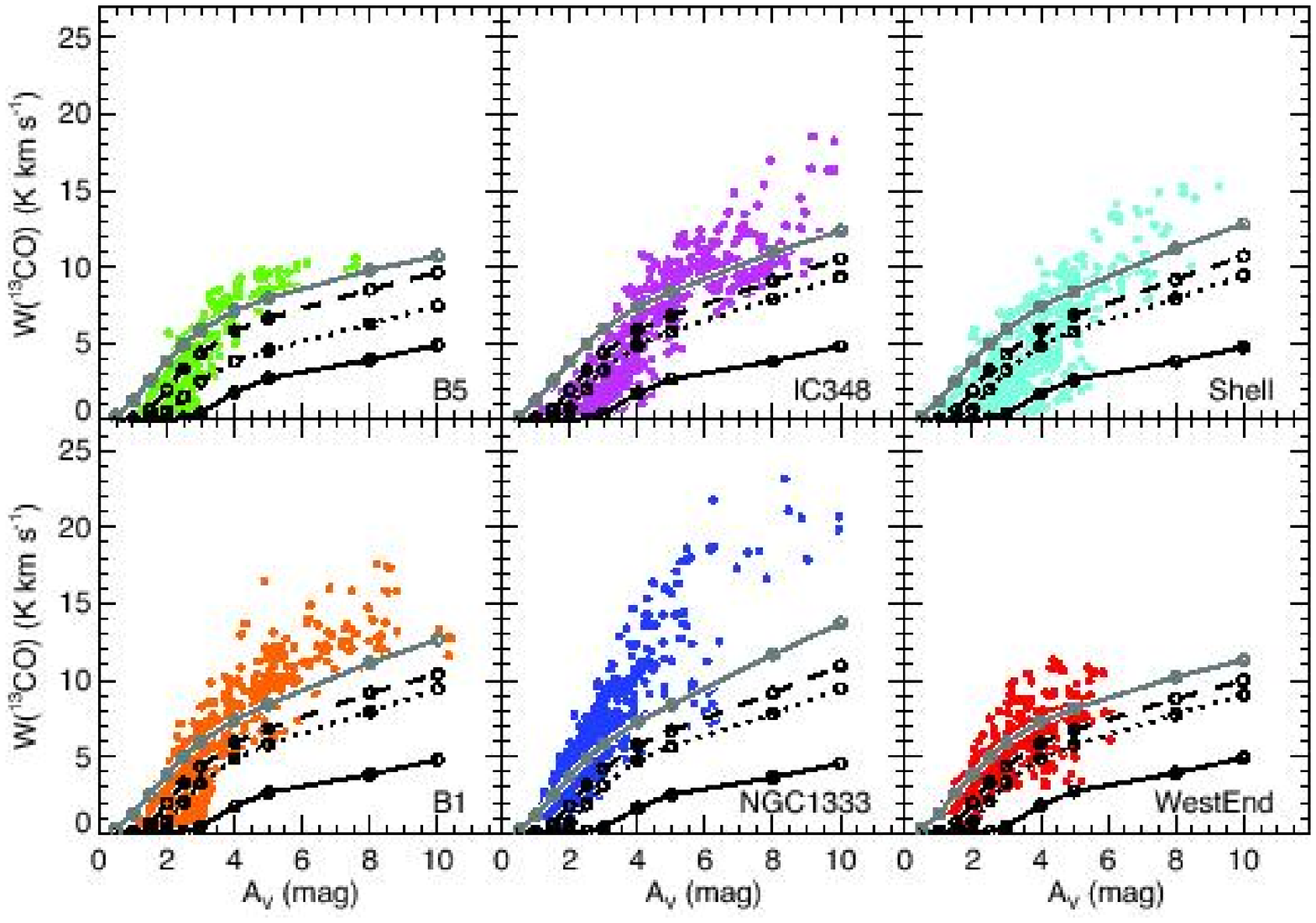}
\plotone{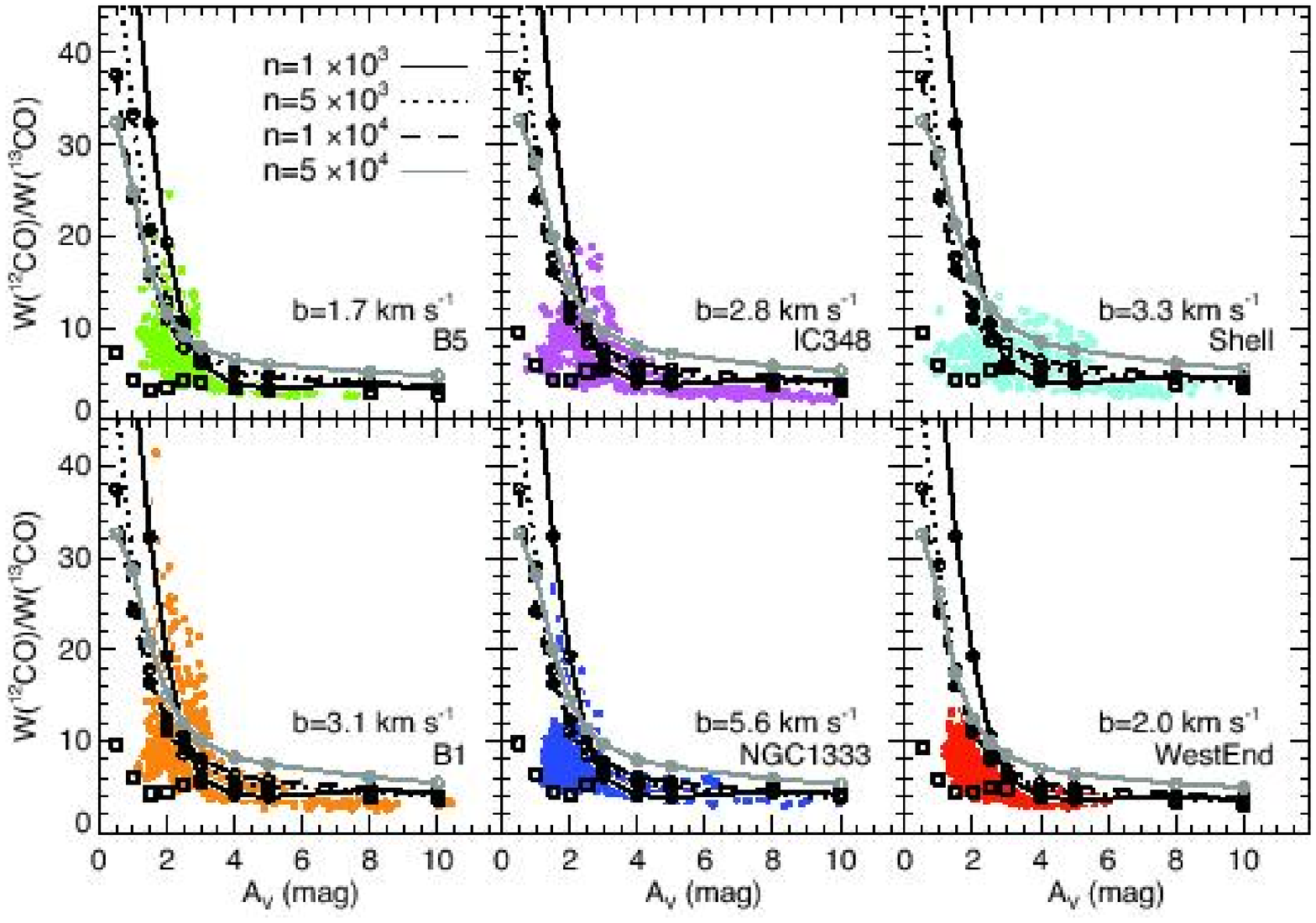}
\caption{\label{pdr} Results of the PDR modeling. Upper, middle 
and bottom panels show the comparison between $W(\mco)$, $W(\mthco)$ 
and the ratio $W(\mco)/W(\mthco)$ with \av, respectively. Black and 
grey lines are derived from PDR models. Unconnected black squares show the ratio 
between \co\ and \thco\ emission predicted by the PDR model with densities 
of $5\times10^3$ and $1\times10^4~\cc$, respectively.}
\end{figure}

The observed values of $W(\mco)$, $W(\mthco)$ and $W(\mco)/W(\mthco)$ 
as a function of \av\ are compared with the models results in Figure~\ref{pdr}. 
To perform this comparison, the PDR code output, $I=\int I_\nu d\nu$, is 
converted using the definition of brightness temperature 
(eq.~\ref{def:T}):
\begin{equation}
W = \int T_R \,dv = \frac{I\,c^2}{2\nu_0^2 k}\frac{c}{\nu_0}
= \frac{c^3}{2\nu_0^3 k} I ~,
\end{equation}
and the ratio can be expressed as 
\begin{equation}
\frac{W(\mco)}{W(\mthco)} = \left(\frac{\nu(\mthco)}{\nu(\mco)}\right)^3 \frac{I(\mco)}{I(\mthco)}
\end{equation}

First, we note that PDR models are reasonably good in reproducing \co\ 
and \thco\ observations, if one allows for variations in densities along different
lines of sight. One exception is NGC1333, 
where the points with ``excess'' in \thco\ emission are associated 
with positions just South of the NGC1333 stellar cluster, where a 
second velocity component is observed. This produces less saturation 
in the \thco\ emission.
Secondly, in the case of \thco, the Doppler parameter (shown in the bottom panels of 
Figure~\ref{pdr})
does not affect the results of PDR models within 10\%, 
whereas for \co, differences of $\le$40\% are present above 2~mag of 
visual extinction. In fact,  the \co\ line is optically thick, thus an 
increase in the line-width produces a significant increase in the line 
brightness because of the optical depth reduction. The insensitivity 
of \thco\ integrated intensity to variations of the turbulent 
line-width leaves the {\it density and radiation field as the two possible 
causes of the observed differences between the regions.}

In Figure~\ref{pdr} we show only the effects of density with $\chi=1$. 
Small variations (within a factor of 3) of the radiation field intensity slightly
shift the curves to higher or lower \av\ by 1~mag, if the radiation field is larger or
smaller, respectively.  The main conclusions for individual regions are listed
below:

\begin{itemize}
\item B5: this is the most quiescent region in the whole Perseus complex. The 
best fit for the $W(\mco)$ is a PDR model of a cloud with  a 
narrow range of density change,  between $5\times10^3$ and $1\times 10^4~\cc$,
values consistent with previous analysis. 
A similar density range is found to 
reproduce  $W(\mthco)$ at $\mav < 4$~mag, whereas
$5\times10^4~\cc$ is more appropriate for the emission at larger 
extinctions (higher density regions are mainly responsible for the \thco\ 
emission at $\mav > 4$~mag). We note that the integrated intensities of the
two CO isotopologues approach (with increasing \av ) the lowest values 
in the whole sample, suggesting
that denser material is hidden to view because of photon trapping in the
narrow range of velocities observed in this region. 

\item IC348: in this cluster-forming region, the density spread is larger 
than in B5, with values ranging between $\rm{few }\times 10^3$ and 
$\simeq 1\times 10^4~\cc$. Similarly, the \thco\ 
at low \av ($<4$~mag) are reproduced by models with $10^3-10^4~\cc$, 
whereas densities larger than $5\times 10^4~\cc$ are needed above 
$4~$mag. However, the points at $\mav>7$~mag and 
$W(\mthco)>12~\kkm$ lie next to the embedded cluster, so that local 
heating and enhanced turbulence are probably increasing the \thco\ brightness 
(the flattening of the $W(\mthco)$~vs.~\av\ curve is in fact less pronounced 
than in the case of B5, suggesting less optical depth). Thus, proto-stellar 
activity is locally affecting the \thco\ emission, but not the \co.

\item Shell: this region shows the largest spread in density for \co\ at all values 
of \av\ from $1\times 10^3$ to about $3\times10^4~\cc$. At $\mav > 5$~mag the data 
groups around two separate values of $W(\mco)$: $\sim 35$ and $55~\kkm$.
The former group is associated with the (outer) shell reported in dust emission by 
\cite{2006ApJ...643..932R}, whereas the latter is located in the inner part of
the shell, 
maybe exposed to a larger radiation field causing more dissociation of CO molecules. 
The \thco\ emission appears similar to 
IC348, with the exception of points located at low $W(\mthco)$ (below $4~\kkm$) and $\mav>4$~mag which are again associated with the inner part of the 
Shell. This also suggests some further destruction process for the \co.

\item B1: the \co\ emission is consistent with material at densities between 
$5\times10^3-3\times10^4~\cc$. Compared with B5, the B1 region shows 
brighter \co\ lines at lower (as well as higher) extinction.  This is probably related
to the larger Doppler parameter of B1. Several data points at low 
\av\ ($<$2~mag) are well reproduced by PDR models of dense clouds. The need of high 
densities at low \av\ also appears in the \thco\ panel. 

\item NGC1333: this is the most active star forming site in the whole Perseus 
cloud and the behaviour of the \co\ and \thco\ integrated intensities as a function
of \av\ is in fact significantly different when compared to the other regions.  First
of all, the densities required in the PDR code to match the data are mostly above 
$10^4~\cc$, both for the \co\ and \thco\ emission. Secondly, the saturation of the
\co\ line becomes evident only at $\mav >6$~mag (unlike $\sim 4$~mag, as in the other
regions). Here, similarly to what is seen in B1, non-thermal motions driven by the 
embedded protostellar cluster are broadening the CO lines, allowing photons from 
deeper in the cloud to escape. The effect is more pronounced than in B1, consistent
with the fact that NGC~1333 has the largest Doppler parameter among the six regions.
We further note that, unlike in IC348, the \co~(1-0) integrated intensity is also
affected by the internal star formation activity, significantly reducing the 
saturation and enhancing the brightness at large \av . Internal motions, likely driven 
by protostellar outflows, are thus
more pronounced in NGC~1333 than in IC348, likely because of the larger star formation 
activity. 

\item \sourcel: this is the only region where no data points are present at 
$\mav >6$~mag, and the \co, as well as the \thco, integrated intensity shows
a large scatter between \av\ of 1 and 6~mag. These two facts are consistent 
with an overall lower 
density and probably clumpy medium, where relatively small high density clumps are 
located along some lines of sight, whereas a significant fraction of the data 
(19\% of points in \co) 
can be reproduced by uniform PDR model clouds with densities below 
$5\times 10^3~\cc$. 
\end{itemize}

In general, model predictions for \thco~(1-0) can only reproduce well the 
observed emission at low extinction ($\mav<3$~mag). The complex structure 
of active star forming regions, in particular density and temperature gradients 
as well as clumpiness along the line of sight (all phenomena not included in the 
PDR code) can of course contribute to the deviations from the uniform PDR models. 
We point out again that the largest Doppler parameters, i.e. the largest amount 
of non--thermal (turbulent?) motions, are present in active regions of star formation, 
so their nature appears to be linked to the current star formation activity and not 
to be part of the initial conditions in the process of star formation. 

In the bottom panel of Fig.~\ref{pdr}, the $W(\mco)/W(\mthco)$ ratio is shown as a 
function of \av\ for the six regions.  The PDR models appear to reproduce well 
the integrated intensity ratio, for a broad range of \av. As we just saw, the \thco\ 
data preferentially trace higher density material than \co\ lines, so the black squares
show the ratio between the \co\ and \thco\ emission as predicted by PDR 
models with $5\times 10^3\cc$ for \co\ and $1\times 10^4\cc$ for \thco\ lines.
One thing to note in these plots
is the large fraction of points at low \av\ and low $W(\mco)/W(\mthco)$ which lie 
below the PDR model curves, in particular for B5, NGC~1333, and \sourcel, but they 
lie above the black squares showing that the PDR model with different densities can
reproduce all the emission. 
Another way to reproduce these data points is by decreasing the interstellar radiation
field by a factor of a few. Alternatively, it is possible that these lines of sight
intercept material where the $^{13}$C carbon is still partially in ionized form, so 
that the reaction 
$^{13}\rm{C}^+ + \mco \rightarrow \rm{^{12}C}^+  + \mthco + \Delta E$ 
(with $\Delta E/k = 37$~K; \citealt{Watson:1977}) can proceed and enhance the 
\thco\ abundance relative to \co.

We finally note here that \cite{Bell:2006} have theoretically investigated 
the variation of the $X$ factor using UCL\_PDR and Meudon PDR codes. They argue 
that variations in $X$ can be due to variations in physical parameters, 
such as the gas density, the radiation field and the turbulence, in 
agreement with our findings. 

\section{Summary and Conclusions}\label{sec-conclusions}
Using the FCRAO \co, \thco\ and \ceio\ data, and a NICER extinction map produced by COMPLETE 
we perform a calibration of the column density estimation using \co, \thco\ and \ceio\ 
emission in Perseus. We report the following results:
\begin{itemize}
\item We find a parameter space, $V_{LSR}(\mco)$--$T_{\rm d}$, in which 
different spatial regions of the Perseus Molecular Cloud Complex also 
cluster, and we designate six regions within the complex (see 
Figures~\ref{vcen-td} and \ref{regions}). We note that the dust temperature 
decreases by about 1~K from North East to South West. 

\item The \co\ data can be modeled with a curve of growth. 
The fit parameters vary between the six regions and this causes  
much of the scatter in the $W(\mco)$~vs.~$W(\mthco)$ plot of the 
whole Perseus Complex. The parameters derived from the fits agree 
with a previous study of a sub-region of Perseus (B5) to within errors 
(see Figure~\ref{i1213}).

\item The $X$ factor,  $X\equiv N(\mhtw)/W(\mco)$, is derived from 
linear fits to the data both for the whole Perseus Complex and for the 
six regions. The \co\ saturates at different intensities in each 
region, depending on the velocity structure of the emission, the 
volume density and radiation field. When the linear fit is done only for the 
unsaturated emission, the $X$ factor is smaller than that derived for 
the Milky Way. However, larger values are obtained (closer to that 
found in the Milky Way) if all the  points are included in the fit (see 
Figure~\ref{co-av} and Table~\ref{table-co-X}).  The most active star
forming region in Perseus (NGC~1333) has the lowest $X$ factor and the
largest \thco\ abundance among the six regions.

\item The gas excitation temperature varies from 4~K to 20~K, it 
increases with \av, and it is typically below the dust temperature at
all \av. This can be explained if a fraction ($\simeq$ 60\%) of the 
\co~(1-0) lines is sub-thermally excited, i.e. if the \co - emitting 
gas has volume densities below $\simeq 3\times 10^3~\cc$. 

\item The column density of \thco\ is derived taking into account the 
effect of optical depth and excitation temperature. We find 
that the threshold extinction above which \thco~(1-0) is detected is 
larger than has previously been reported.  However, the fractional 
abundances (w.r.t. \htw\ molecules) are in agreement with previous 
determinations. The difference with previous works is  due to
the superior zero-point calibration and larger dynamic range of the
NICER extinction map, as compared to those derived from optical star 
counting (see Figure~\ref{13co-av} and Table~\ref{table-cal}).

\item \thco\ abundance variations between the regions do not correlate
with the extinction threshold \avo , suggesting that the main cause of the 
variation is likely due to the chemical/physical properties of shielded 
molecular material deeper into the cloud.  The \co~(1-0) and  \thco~(1-0)
lines saturates at $\mav > 4, 5$~mag, respectively, whereas  
\ceio~(1-0) line do not show signs of saturation up to the largest 
\av\ probed by our data (10~mag). 

\item Using the Meudon PDR code we find that the observed variations
among the different regions can be explained with variations in physical
parameters, in particular the volume density and internal motions. 
Large Doppler parameters imply large values of the CO integrated intensities
(as expected for very optically thick lines) and are typically found in
active star forming regions (the largest values of the Doppler parameter
and $W(\mco)$ being associated with NGC~1333, the most active site of star 
formation in Perseus).  On the other hand, quiescent regions such as B5 
appears less bright in CO and only show a narrow range of CO integrated 
intensities as a function of \av.  This is likely due to the fact that 
the photons emitted from the higher density regions located deep into 
the cloud have similar velocities relative to the outer cloud envelope 
traced by \co, so that they are more easily absorbed. Thus, turbulent 
(or, more generally, non-thermal)  motions appear to be a by-product 
of star formation, more than part of 
the initial conditions in the star formation process. 
\end{itemize}

This work has shown that local variations in physical conditions 
significantly affect the relation between CO-isotopologue emission and 
\av , contributing to the observed scatter. The use of a standard $X$ 
factor, $1.8\times 10^{20}\,\cmkm$, produces an overestimation of the 
cloud's mass by $\sim 45\%$ when compared to the mass derived from 
the extinction map, while the lower limit for the mass derived using 
the linear fit to the unsaturated points underestimates the mass by a 
$\sim 15\%$. The $X$ factor (as well as the 
\thco\ fractional abundance) depends on the star 
formation activity, with lower values associated with the more 
active (and turbulent) regions.  
Extinctions measured by using \thco\ and previous 
conversions from the literature are typically underestimated by $\sim 
0.8$~mag, so that more shielding is needed to produce the observed \thco\ 
compared to previous findings.

\acknowledgments
JEP is supported by the National Science Foundation through grant 
\#AF002 from the Association of Universities for Research in 
Astronomy, Inc., under NSF cooperative agreement AST-9613615 and by 
Fundaci\'on Andes under project No. C-13442. This material is based 
upon work supported by the National Science Foundation under Grant 
No. AST-0407172.  PC acknowledges support by the Italian Ministry 
of Research and University within a PRIN project.

\appendix

\section{Effect of $\chi$ and $\zeta$ variation}

\begin{figure}
\centering
\epsscale{0.8}
\plotone{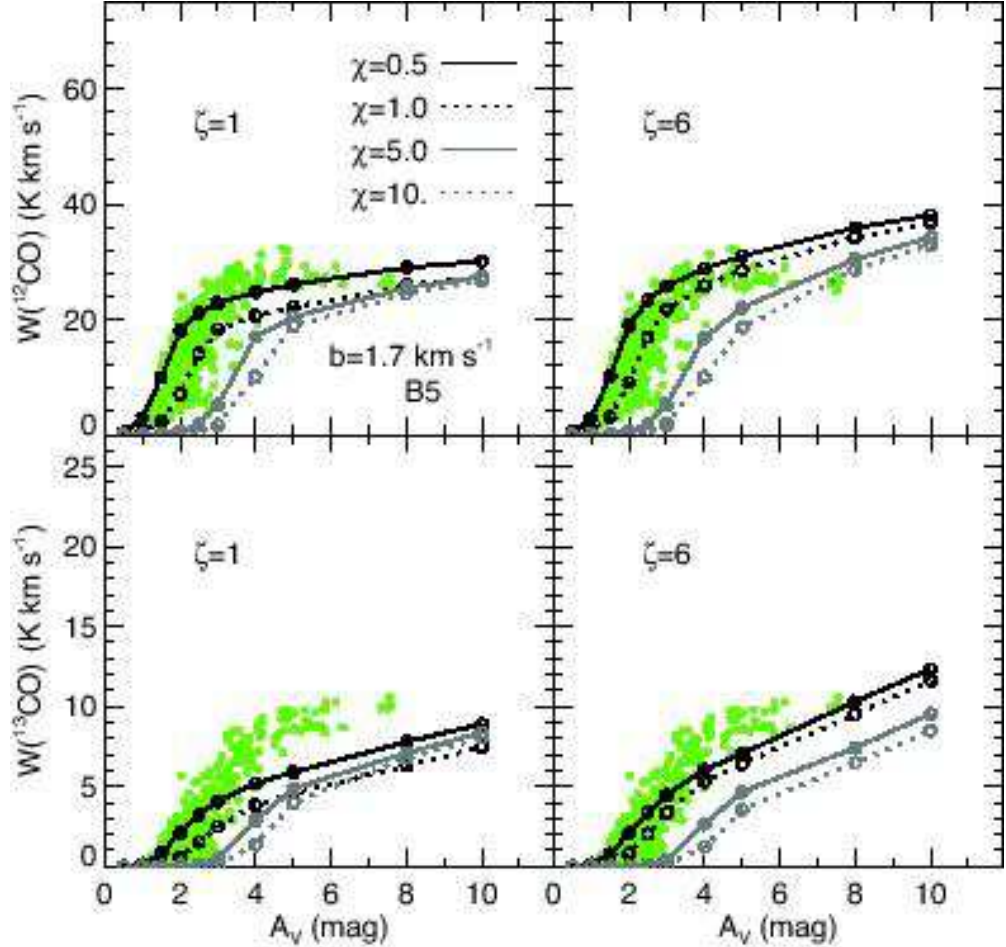}
\caption{PDR results for the particular case of B5 and different values of $\chi$ and 
$\zeta$ (in units of $10^{-17}~\rm{s}^{-1}$). 
Black solid, black dotted, gray solid and gray dotted lines show the PDR model results 
for $\chi=0.5,1.0,5.0~{\rm and}~10$ times the standard Draine's radiation field, respectively.
Variations in $\chi$ mainly affect the 
low \av\ emission, whereas different $\zeta$ values slightly change the emission values 
at  $\mav > 4$~mag. \label{fig_app}}
\end{figure}

In Sect.~\ref{sec-pdr} we have explored how changes in volume density and Doppler parameter 
affect the \co~(1-0) and \thco~(1-0) integrated intensities ($W(\mco)$, $W(\mthco)$) predicted 
by the Meudon PDR code. Here we show the effects of variations in the interstellar radiation 
field intensity, in units of Draine's field ($\chi$), and the cosmic-ray ionization rate ($\zeta$) on 
$W(\mco)$ and $W(\mthco)$.  Fig.~\ref{fig_app} shows the results of this parameter space 
exploration in the particular case of B5  and 
volume density of $5\times 10^{3}~\cc$ (similar results apply to the other regions and different
densities).  The upper panels display the model results for $\chi$ values of 0.5, 1, 5 and 10: the main 
change is visible at $\mav < 4$~mag, with a shift of the threshold extinction for \co\ and \thco\ 
emission from about 1 to 3~mag for an increase of $\chi$ from 0.5 to 10, respectively.  

The cosmic-ray ionization rate used in the PDR models described in Sect.~\ref{sec-pdr} is
$\zeta = 1\times 10^{-17}$~s$^{-1}$.  This value is quite uncertain and \cite{Dalgarno:2006}
suggests a higher rate of $6\times 10^{-17}~\rm{s}^{-1}$ for molecular clouds 
\citep[see also][]{vardertak:2000}.  In the bottom panels of Fig.~\ref{fig_app}, 
we show the predicted $W(\mco)$ and $W(\mthco)$ curves for $\zeta = 
6\times 10^{-17}~\rm{s}^{-1}$.  The larger $\zeta$ value does not affect the 
CO emission at $\mav \leq 3$~mag, and only changes the integrated intensities by 
about 30\%.  Thus, the effect is not large enough to explain the largest $W(\mthco)$
values.

\begin{deluxetable}{lccc}
\tablecolumns{4}
\tablecaption{Clusters Regions Removed\label{tab:cluster}}
\tablehead{
\colhead{Region} & \colhead{Center R.A.} & \colhead{Center Decl.} & \colhead{Box Size}\\
\colhead{} & \colhead{(deg)} & \colhead{(deg)} & \colhead{(deg$^{2}$)}
}
\startdata
IC 348 		& 56.088 & 32.171 & 0.480$\times$0.426\\
NGC 1333 	& 52.212 & 31.483 & 0.397$\times$0.496
\enddata
\end{deluxetable}

\clearpage

\begin{deluxetable}{lcccccccccc}
\rotate
\centering
\tabletypesize{\scriptsize}
\tablecolumns{11}
\tablewidth{0pt}
\tablecaption{Typical Properties of the Regions from Average Spectra\label{tab:prop}}
\tablehead{ 
\colhead{Region} 
& \colhead{$W(\mco)$} & \colhead{$T_{max}(\mco)$} & 
\colhead{$V_{LSR}(\mco)$} & \colhead{$\sigma_V(\mco)$} & 
\colhead{$W(\mthco)$} & \colhead{$T_{max}(\mthco)$} & 
\colhead{$V_{LSR}(\mthco)$} & \colhead{$\sigma_V(\mthco)$} & 
\colhead{$T_{ex}$} & \colhead{$\tau(\mthco)$} 
\\
\colhead{} 
& \colhead{(K\ km\ s$^{-1}$)} 
& \colhead{(K)}
& \colhead{(km\ s$^{-1}$)} 
& \colhead{(km\ s$^{-1}$)} 
& \colhead{(K\ km\ s$^{-1}$)} 
& \colhead{(K)}
& \colhead{(km\ s$^{-1}$)} 
& \colhead{(km\ s$^{-1}$)} 
& \colhead{(K)} 
& \colhead{} 
}
\startdata
B5		&    8&   3.2	&   9.84	&   0.90&   1.6&   0.99&   9.99&   0.64&  12&   0.31\\
IC348	&    9&   3.0	&   9.01	&   1.18&   2.2&   0.93&   8.99&   0.95&  12&   0.35\\
Shell		&  10&   2.6	&   8.73	&   1.49&   2.0&   0.58&   8.72&   1.30&  11&   0.28\\
B1		&  12&   2.6	&   6.71	&   1.55&   1.6&   0.99&   6.83&   0.99&  11&   0.33\\
NGC1333	&  15&   3.0	&   6.68	&   1.85&   2.6&   0.73&   7.06&   1.38&  11&   0.31\\
\sourcel	&  10&   2.2	&   4.20	&   1.96&   2.0&   0.65&   4.58&   1.19&    9&   0.44\\
Perseus	&  11&   1.7	&   7.26	&   2.65&   2.2&   0.39&   7.67&   2.15&  11&   0.34
\enddata
\tablecomments{
$W\equiv$ integrated intensity.
$T_{max}\equiv$ peak brightness temperature.
$V_{LSR}\equiv$ centroid velocity.
$\sigma_{V}\equiv$ velocity dispersion ($=\Delta v/\sqrt{8\ln{2}}$, where $\Delta v$ is the full width at half maximum).
$T_{ex}\equiv$ excitation temperature derived from \co.
$\tau\equiv$ \thco\ optical depth.
The 1-$\sigma$ uncertainty for integrated intensity, peak brightness and excitation temperature is
estimated between 15 and 30\%.
}
\end{deluxetable}

\begin{deluxetable}{lcc}
\tablecolumns{3}
\tablecaption{Parameters for Growth Curve Fit \label{table-growth}}
\tablehead{ \colhead{Region} & \colhead{$a$} & \colhead{$T_R\,b$} \\
\colhead{} & \colhead{(${\rm K^{-1}\,km^{-1}\,s}$)} & \colhead{($\kkm$)}}
\startdata
B5      	&        0.61$\pm$0.03 	& 20.5$\pm$0.5 \\
IC348   	&      0.246$\pm$0.007 	& 33.5$\pm$0.6 \\
Shell   	&      0.223$\pm$0.009 	&    40$\pm$1 \\
B1      	&        0.33$\pm$0.01 	& 36.8$\pm$0.7 \\
NGC1333	&      0.130$\pm$0.006 	&    67$\pm$2 \\
\sourcel	&        0.44$\pm$0.03 	& 24.7$\pm$0.9 \\
Perseus 	&      0.260$\pm$0.004 	& 35.9$\pm$0.4 
\enddata
\end{deluxetable}

\begin{deluxetable}{lccc}
\tabletypesize{\footnotesize}
\tablecolumns{4}
\tablecaption{Linear fits  to \co\label{table-co-X}}
\tablehead{ 
\colhead{Region} &  \colhead{$X/10^{20}$} & \colhead{$A_{V12}$} & \colhead{$X_2/10^{20}$} \\
\colhead{} 		& \colhead{($\cmkm$)} & \colhead{(mag)} 	& \colhead{($\cmkm$)} 
}
\startdata
\cutinhead{Fit performed to the whole dataset}
B5       	&         2$\pm$1 	&       0.0$\pm$0.2 	&           1.5$\pm$0.1 \\
IC348    	&         3$\pm$2 	&     0.51$\pm$0.08 	&        1.61$\pm$0.07 \\
Shell    	&         2$\pm$1 	&       1.0$\pm$0.1 	&        1.09$\pm$0.08 \\
B1       	&      1.4$\pm$0.8 	&      -0.6$\pm$0.1 	&        1.38$\pm$0.09 \\
NGC1333	&      0.9$\pm$0.3 	&      -0.5$\pm$0.1 	&        0.93$\pm$0.07 \\
\sourcel 	&      1.2$\pm$0.5 	&      -1.8$\pm$0.3 	&          1.8$\pm$0.4   \\
Perseus  	&         2$\pm$1 	&    -0.33$\pm$0.07	&        1.38$\pm$0.05 \\
\cutinhead{Fit performed to points where $\mav < 4$}
B5       	&         2$\pm$1 	&      0.97$\pm$0.06	&     0.88$\pm$0.09 \\
IC348    	&         3$\pm$2 	&      1.60$\pm$0.04	&     0.76$\pm$0.06 \\
Shell    	&         2$\pm$2 	&        1.5$\pm$0.1 	&       0.8$\pm$0.1 \\
B1       	&      1.4$\pm$0.9 	&      1.18$\pm$0.04	&     0.56$\pm$0.04 \\
NGC1333	&      0.9$\pm$0.3 	&      0.60$\pm$0.06 	&     0.57$\pm$0.06 \\
\sourcel 	&      1.1$\pm$0.4 	&       -0.4$\pm$0.2 	&       1.2$\pm$0.3 \\
Perseus 	&         2$\pm$1 	&      0.92$\pm$0.04 	&     0.72$\pm$0.04\\
\cutinhead{Previous works}
Galaxy (1) 	&       2.8$\pm$0.4 	&     \nodata	 	&    \nodata        \\
$\rho$-Oph (2) 	&       1.8$\pm$0.1 	&     \nodata	 	&    \nodata        \\
Galaxy (3) 	&       1.8$\pm$0.3 	&     \nodata	 	&    \nodata        \\
Pipe (4)  		&       \nodata 		&     2.02$\pm$0.02 	&    1.06$\pm$0.02   
\enddata
\tablerefs{
(1) \cite{Bloemen:1986};
(3) \cite{FLW82};
(3) \cite{Dame:2001};
(4) \cite{lombardi:pipe};
}
\end{deluxetable}

\begin{deluxetable}{lccc}
\tablecolumns{4}
\tablecaption{Parameters for \co\ Fit \label{table-co}}
\tablehead{ \colhead{Region} & \colhead{$W_0$} & \colhead{k} & \colhead{$A_{k12}$} \\
\colhead{} & \colhead{($\kkm$)}  & \colhead{(mag$^{-1}$)} & \colhead{(mag)} 
}
\startdata
B5       	&     30.92 &    0.553 &     1.063 \\
IC348    	&     39.27 &    0.350 &     1.374 \\
Shell    	&     73.31 &    0.139 &     0.851 \\
B1       	&     43.08 &    0.691 &     1.309 \\
NGC1333	&     67.12 &    0.374 &     0.748 \\
\sourcel 	&     30.45 &    0.529 &   -0.160 \\
Perseus	&     42.29 &    0.367 &     0.580
\enddata
\end{deluxetable}

\begin{deluxetable}{lccc}
\tablecolumns{4}
\tablecaption{Results of Linear Fit to $W(\mthco)$\label{table-w13}}
\tablehead{
\colhead{Region} &
\colhead{$A_{W13}$} & 
\colhead{$B_{13}$} & 
\colhead{Reference}\\
\colhead{} & 
\colhead{(mag)} & 
\colhead{(mag\,K$^{-1}$\,km$^{-1}$\,s)} & 
\colhead{}
}
\startdata
B5      	&  1.56$\pm$0.03 &  0.323$\pm$0.009 	& 1\\
IC348   	&  1.99$\pm$0.03 &  0.36$\pm$0.01 	& 1\\
Shell   	&  1.90$\pm$0.06 &  0.40$\pm$0.02 	& 1\\
B1      	&  1.64$\pm$0.03 &  0.296$\pm$0.009 	& 1\\
NGC1333	&  1.19$\pm$0.03 &  0.26$\pm$0.01 	& 1\\
Westend 	&  0.75$\pm$0.06 &  0.44$\pm$0.02 	& 1\\
Perseus 	&  1.46$\pm$0.02 &  0.345$\pm$0.006  	& 1\\
\cutinhead{Previous works}
IC~5146	&   -2.6$\pm$0.3    &     1.4$\pm$0.1    &  2\\
B5     	&   0.54$\pm$0.13 &   0.39$\pm$0.02   &  3
\enddata
\tablerefs{
(1) This work, linear fit for $\mav<5$; 
(2) \cite{Lada:1994}, linear fit for $\mav<5$;
(3) \cite{1989ApJ...337..355L}
}
\end{deluxetable}

\begin{deluxetable}{lcccc}
\tablecolumns{5}
\tablecaption{Results of Linear Fit to $N(\mthco)$\label{table-cal}}
\tablehead{ \colhead{Region} & \colhead{\avo} & \colhead{c}& \colhead{[\htw/\thco]}
& \colhead{Reference} \\
\colhead{} & \colhead{(mag)} & \colhead{(mag\ cm$^{-2}$)} & \colhead{$\times 10^{5}$}
& \colhead{}}
\startdata
B5      		&   1.62$\pm$0.04 	&   (4.0$\pm$0.2)$\times 10^{-16}$ 		& 3.8$\pm$0.2 		& 1,a\\ 
IC348   		&   2.15$\pm$0.04 	&   (4.4$\pm$0.1)$\times 10^{-16}$ 		& 4.1$\pm$0.1 		& 1,a\\ 
Shell   		&   2.23$\pm$0.05 	&   (4.3$\pm$0.1)$\times 10^{-16}$ 		& 4.1$\pm$0.1 		& 1,a\\ 
B1      		&   1.68$\pm$0.04 	&   (4.0$\pm$0.1)$\times 10^{-16}$ 		& 3.8$\pm$0.1		& 1,a\\ 
NGC1333		&   1.44$\pm$0.04 	&   (3.0$\pm$0.1)$\times 10^{-16}$ 		& 2.8$\pm$0.1 		& 1,a\\ 
Westend 		&   1.19$\pm$0.05 	&   (5.2$\pm$0.2)$\times 10^{-16}$ 		& 4.9$\pm$0.2 		& 1,a\\ 
Perseus 		&   1.67$\pm$0.02 	&   (4.24$\pm$0.07)$\times 10^{-16}$ 	& 3.98$\pm$0.07 	& 1,a\\ 
\cutinhead{Previous works}
Perseus~B5   	& 0.5$\pm$0.1   	&   (3.8$\pm$0.2)$\times 10^{-16}$         	& 3.6$\pm$0.2         	& 2,b\\
Perseus         	& 0.8$\pm$0.4        	&   (4.0$\pm$0.8)$\times 10^{-16}$         	& 3.8$\pm$0.8   	& 3,b\\
L1495           	& 0.3$\pm$0.3        	&    (4.5$\pm$0.6)$\times 10^{-16}$        	& 4.2$\pm$0.6           & 4,b\\ L1517           	& 0.3$\pm$0.5        	&    (6$\pm$2)$\times 10^{-16}$           	&    5$\pm$1               & 4,b\\ $\rho$-Oph      	& 1.6$\pm$0.3         	&    (3.7$\pm$0.4)$\times 10^{-16}$       	& 3.5$\pm$0.4           & 5,c\\
Taurus          	& 1.0$\pm$0.2         	&    (7.1$\pm$0.7)$\times 10^{-16}$  	& 6.7$\pm$0.7           & 5,c
\enddata
\tablecomments{(a) \av\ derived from NIR colors, (b) \av\ derived from star counting; (c) \av\ derived using spectra.}
\tablerefs{
(1) This work;
(2) \cite{1989ApJ...337..355L};
(3) \cite{1986A&A...166..283B};
(4) \cite{1986A&A...164..349D};
(5) \cite{FLW82}  }
\end{deluxetable}

\begin{deluxetable}{lccc}
\tablecolumns{4}
\tablecaption{Results of Linear Fit to $W(\mceio)$\label{table-w18}}
\tablehead{ \colhead{Region} & \colhead{$A_{W18}$} & \colhead{$B_{18}$} & \colhead{Reference}\\
\colhead{} & \colhead{(mag)} & \colhead{(mag\,K$^{-1}$\,km$^{-1}$\,s)} & \colhead{}
}
\startdata
IC348   	&    2.1$\pm$0.4 	&    5$\pm$11 		& 1\\
Shell   	&    2.7$\pm$0.3 	&    4$\pm$7 		& 1\\
B1      	&    2.1$\pm$0.1 	&    2.5$\pm$0.9 	& 1\\
NGC1333	&    1.7$\pm$0.2 	&    3$\pm$2 		& 1\\
Westend 	&    1.6$\pm$0.3 	&    3$\pm$5 		& 1\\
Perseus 	&    2.4$\pm$0.1 	&    2.9$\pm$0.9  	& 1\\
\cutinhead{Previous works}
IC~5146	&   -0.7$\pm$0.3     	&    10$\pm$1 		& 2\\
B5     	&  1.40$\pm$0.22 	&   1.8$\pm$0.13	&  3
\enddata
\tablerefs{
(1) This work; 
(2) \cite{Lada:1994}, linear fit for $\mav<15$;
(3) \cite{1989ApJ...337..355L}
}
\end{deluxetable}

\begin{deluxetable}{lc}
\tablecolumns{2}
\tablecaption{Initial chemical abundances with respect to total Hydrogen \label{table:abun}}
\tablehead{
\colhead{Element} & \colhead{Abundance}
}
\startdata
He 		& $0.1$\\
C$^+$ 	& $1.307\times 10^{-4}$\\
$^{13}$C$^+$ 	& $1.633\times 10^{-6}$\\
N 		& $2.14\times 10^{-5}$\\
O 		& $1.76\times 10^{-4}$\\
S$^+$ 	& $8.00\times 10^{-6}$\\
Fe$^+$ 	& $3.00\times 10^{-7}$
\enddata
\end{deluxetable}

\clearpage

\begin{thebibliography}

\bibitem[{{Akritas} \& {Bershady}(1996)}]{1996ApJ...470..706A}
{Akritas}, M.~G., \& {Bershady}, M.~A. 1996, \apj, 470, 706

\bibitem[{{Alves} {et~al.}(2007){Alves}, {Lombardi}, \& {Lada}}]{pers-NIR}
{Alves}, J., {Lombardi}, M., \& {Lada}, C.~J. 2007, In prep.

\bibitem[{{Arce} \& {Goodman}(1999)}]{1999ApJ...517..264A}
{Arce}, H.~G., \& {Goodman}, A.~A. 1999, \apj, 517, 264

\bibitem[{{Bachiller} \& {Cernicharo}(1986)}]{1986A&A...166..283B}
{Bachiller}, R., \& {Cernicharo}, J. 1986, \aap, 166, 283

\bibitem[{{Bell} {et~al.}(2006){Bell}, {Roueff}, {Viti}, \&
  {Williams}}]{Bell:2006}
{Bell}, T.~A., {Roueff}, E., {Viti}, S., \& {Williams}, D.~A. 2006, \mnras,
  371, 1865

\bibitem[{{Bensch}(2006)}]{Bensch:2006}
{Bensch}, F. 2006, \aap, 448, 1043

\bibitem[{{Bloemen} {et~al.}(1986){Bloemen}, {Strong}, {Mayer-Hasselwander},
  {Blitz}, {Cohen}, {Dame}, {Grabelsky}, {Thaddeus}, {Hermsen}, \&
  {Lebrun}}]{Bloemen:1986}
{Bloemen}, J.~B.~G.~M., {Strong}, A.~W., {Mayer-Hasselwander}, H.~A., {Blitz},
  L., {Cohen}, R.~S., {Dame}, T.~M., {Grabelsky}, D.~A., {Thaddeus}, P.,
  {Hermsen}, W., \& {Lebrun}, F. 1986, \aap, 154, 25

\bibitem[{{Bohlin} {et~al.}(1978){Bohlin}, {Savage}, \& {Drake}}]{bohlin:1978}
{Bohlin}, R.~C., {Savage}, B.~D., \& {Drake}, J.~F. 1978, \apj, 224, 132

\bibitem[{{Cambr{\'e}sy}(1999)}]{Cambresy:1999}
{Cambr{\'e}sy}, L. 1999, \aap, 345, 965

\bibitem[{{Cernicharo} \& {Bachiller}(1984)}]{1984A&AS...58..327C}
{Cernicharo}, J., \& {Bachiller}, R. 1984, \aaps, 58, 327

\bibitem[{{Combes}(1991)}]{Combes:ARAA}
{Combes}, F. 1991, \araa, 29, 195

\bibitem[{{Dalgarno}(2006)}]{Dalgarno:2006}
{Dalgarno}, A. 2006, Proceedings of the National Academy of Science, 103, 12269

\bibitem[{{Dame} {et~al.}(2001){Dame}, {Hartmann}, \& {Thaddeus}}]{Dame:2001}
{Dame}, T.~M., {Hartmann}, D., \& {Thaddeus}, P. 2001, \apj, 547, 792

\bibitem[{{de Vries} {et~al.}(1987){de Vries}, {Thaddeus}, \&
  {Heithausen}}]{deVries:Ursa}
{de Vries}, H.~W., {Thaddeus}, P., \& {Heithausen}, A. 1987, \apj, 319, 723

\bibitem[{{Draine}(1978)}]{Draine:1978}
{Draine}, B.~T. 1978, \apjs, 36, 595

\bibitem[{{Draine}(2003)}]{Draine:Review}
---. 2003, \araa, 41, 241

\bibitem[{{Duvert} {et~al.}(1986){Duvert}, {Cernicharo}, \&
  {Baudry}}]{1986A&A...164..349D}
{Duvert}, G., {Cernicharo}, J., \& {Baudry}, A. 1986, \aap, 164, 349

\bibitem[{{Elias}(1978)}]{Elias:Oph}
{Elias}, J.~H. 1978, \apj, 224, 453

\bibitem[{{Frerking} {et~al.}(1982){Frerking}, {Langer}, \& {Wilson}}]{FLW82}
{Frerking}, M.~A., {Langer}, W.~D., \& {Wilson}, R.~W. 1982, \apj, 262, 590

\bibitem[{{Goldsmith}(2001)}]{Goldsmith:2001}
{Goldsmith}, P.~F. 2001, \apj, 557, 736

\bibitem[{{Goodman} {et~al.}(2007){Goodman}, {Pineda},  \&
  {Schnee}}]{alyssa-lognormal}
{Goodman}, A.~A., {Pineda}, J.~E., \& {Schnee}, S.~L. 2007, In
  prep.

\bibitem[{{Hatchell} \& {van der Tak}(2003)}]{Hatchell:2003}
{Hatchell}, J., \& {van der Tak}, F.~F.~S. 2003, \aap, 409, 589

\bibitem[{{Hily-Blant} {et~al.}(2005-1){Hily-Blant}, {Pety}, \& S.}]{class90}
{Hily-Blant}, P., {Pety}, J., \& S., G. 2005-1, CLASS evolution: I. Improved
  OFT support, Tech. rep., IRAM

\bibitem[{{Lada} {et~al.}(1994){Lada}, {Lada}, {Clemens}, \&
  {Bally}}]{Lada:1994}
{Lada}, C.~J., {Lada}, E.~A., {Clemens}, D.~P., \& {Bally}, J. 1994, \apj, 429,
  694

\bibitem[{{Langer} {et~al.}(1989){Langer}, {Wilson}, {Goldsmith}, \&
  {Beichman}}]{1989ApJ...337..355L}
{Langer}, W.~D., {Wilson}, R.~W., {Goldsmith}, P.~F., \& {Beichman}, C.~A.
  1989, \apj, 337, 355

\bibitem[{{Le Petit} {et~al.}(2006){Le Petit}, {Nehm{\'e}}, {Le Bourlot}, \&
  {Roueff}}]{Meudon:PDR}
{Le Petit}, F., {Nehm{\'e}}, C., {Le Bourlot}, J., \& {Roueff}, E. 2006, \apjs,
  164, 506

\bibitem[{{Lee} {et~al.}(1998){Lee}, {Roueff}, {Pineau des Forets},
  {Shalabiea}, {Terzieva}, \& {Herbst}}]{Lee:1998}
{Lee}, H.-H., {Roueff}, E., {Pineau des Forets}, G., {Shalabiea}, O.~M.,
  {Terzieva}, R., \& {Herbst}, E. 1998, \aap, 334, 1047

\bibitem[{{Lombardi} \& {Alves}(2001)}]{NICER:Lombardi-Alves}
{Lombardi}, M., \& {Alves}, J. 2001, \aap, 377, 1023

\bibitem[{{Lombardi} {et~al.}(2006){Lombardi}, {Alves}, \&
  {Lada}}]{lombardi:pipe}
{Lombardi}, M., {Alves}, J., \& {Lada}, C.~J. 2006, \aap, 454, 781

\bibitem[{{Miville-Desch{\^e}nes} \& {Lagache}(2005)}]{Miville05}
{Miville-Desch{\^e}nes}, M.-A., \& {Lagache}, G. 2005, \apjs, 157, 302

\bibitem[{{Padoan} {et~al.}(1999){Padoan}, {Bally}, {Billawala}, {Juvela}, \&
  {Nordlund}}]{Padoan:Perseus}
{Padoan}, P., {Bally}, J., {Billawala}, Y., {Juvela}, M., \& {Nordlund}, {\AA}.
  1999, \apj, 525, 318

\bibitem[{{Ridge} {et~al.}(2006{\natexlab{a}}){Ridge}, {Di Francesco}, {Kirk},
  {Li}, {Goodman}, {Alves}, {Arce}, {Borkin}, {Caselli}, {Foster}, {Heyer},
  {Johnstone}, {Kosslyn}, {Lombardi}, {Pineda}, {Schnee}, \&
  {Tafalla}}]{COMPLETE-I}
{Ridge}, N.~A., {Di Francesco}, J., {Kirk}, H., {Li}, D., {Goodman}, A.~A.,
  {Alves}, J.~F., {Arce}, H.~G., {Borkin}, M.~A., {Caselli}, P., {Foster},
  J.~B., {Heyer}, M.~H., {Johnstone}, D., {Kosslyn}, D.~A., {Lombardi}, M.,
  {Pineda}, J.~E., {Schnee}, S.~L., \& {Tafalla}, M. 2006{\natexlab{a}}, \aj,
  131, 2921

\bibitem[{{Ridge} {et~al.}(2006{\natexlab{b}}){Ridge}, {Schnee}, {Goodman}, \&
  {Foster}}]{2006ApJ...643..932R}
{Ridge}, N.~A., {Schnee}, S.~L., {Goodman}, A.~A., \& {Foster}, J.~B.
  2006{\natexlab{b}}, \apj, 643, 932

\bibitem[{{Rieke} \& {Lebofsky}(1985)}]{Rieke:1985}
{Rieke}, G.~H., \& {Lebofsky}, M.~J. 1985, \apj, 288, 618

\bibitem[R{\"o}llig et al.(2007)]{PDR:comparison} R{\"o}llig, M., et 
al.\ 2007, \aap, 467, 187 

\bibitem[{{Rohlfs} \& {Wilson}(1996)}]{ToolsRA}
{Rohlfs}, K., \& {Wilson}, T.~L. 1996, {Tools of Radio Astronomy} (Tools of
  Radio Astronomy, XVI, 423 pp.~127 figs., 20 tabs..~ Springer-Verlag Berlin
  Heidelberg New York.~Also Astronomy and Astrophysics Library)

\bibitem[{{Schnee} {et~al.}(2006){Schnee}, {Bethell}, \&
  {Goodman}}]{Schnee:Bethell:Goodman}
{Schnee}, S., {Bethell}, T., \& {Goodman}, A. 2006, \apjl, 640, L47

\bibitem[{{Schnee} {et~al.}(2008){Schnee}, {Li}, {Goodman}, \&
  {Sargent}}]{Schnee:spitzer}
{Schnee}, S.~L., {Li}, J.~G., {Goodman}, A.~A., \& {Sargent}, A.~I. 2008, In
  prep.

\bibitem[{{Schnee} {et~al.}(2005){Schnee}, {Ridge}, {Goodman}, \&
  {Li}}]{scott-temp}
{Schnee}, S.~L., {Ridge}, N.~A., {Goodman}, A.~A., \& {Li}, J.~G. 2005, \apj,
  634, 442

\bibitem[Solomon et al.(1987)]{Solomon:relations} Solomon, P.~M., Rivolo, 
A.~R., Barrett, J., \& Yahil, A.\ 1987, \apj, 319, 730

\bibitem[{{Spitzer}(1968)}]{spitzer:1968}
{Spitzer}, L. 1968, {Diffuse matter in space} (New York: Interscience
  Publication, 1968)

\bibitem[{{Spitzer}(1978)}]{spitzer:1978}
---. 1978, {Physical processes in the interstellar medium} (New York
  Wiley-Interscience, 1978.~333 p.)

\bibitem[{{Strong} \& {Mattox}(1996)}]{Strong-Mattox:1996}
{Strong}, A.~W., \& {Mattox}, J.~R. 1996, \aap, 308, L21

\bibitem[{{Tielens}(2005)}]{Tielens:book}
{Tielens}, A.~G.~G.~M. 2005, {The Physics and Chemistry of the Interstellar
  Medium} (The Physics and Chemistry of the Interstellar Medium, by
  A.~G.~G.~M.~Tielens, pp.~.~ISBN 0521826349.~Cambridge, UK: Cambridge
  University Press, 2005.)

\bibitem[{{van der Tak} \& {van Dishoeck}(2000)}]{vardertak:2000}
{van der Tak}, F.~F.~S., \& {van Dishoeck}, E.~F. 2000, \aap, 358, L79

\bibitem[{{Watson}(1977)}]{Watson:1977}
{Watson}, W.~D. 1977, in ASSL Vol. 67: CNO Isotopes in Astrophysics, ed.
  J.~{Audouze}, 105--114

\bibitem[{{Wilson} {et~al.}(1970){Wilson}, {Jefferts}, \&
  {Penzias}}]{Wilson:CO}
{Wilson}, R.~W., {Jefferts}, K.~B., \& {Penzias}, A.~A. 1970, \apjl, 161, L43+

\bibitem[{{Young} {et~al.}(1982){Young}, {Goldsmith}, {Langer}, {Wilson}, \&
  {Carlson}}]{Young:B5}
{Young}, J.~S., {Goldsmith}, P.~F., {Langer}, W.~D., {Wilson}, R.~W., \&
  {Carlson}, E.~R. 1982, \apj, 261, 513

\bibitem[{{Young} \& {Scoville}(1982)}]{Young:1982}
{Young}, J.~S., \& {Scoville}, N. 1982, \apj, 258, 467

\end{thebibliography}
\end{document}